\documentclass[letterpaper]{aa501}

\usepackage{rotating} 
\usepackage{graphics}
\usepackage{graphicx}

\newfont{\tf}{cmr10 scaled 580}
\newfont{\ttf}{cmtt10 scaled 580} 
\newcommand{\omm}{$\scriptstyle -$}
\newcommand{\omp}{$\scriptstyle +$}

\def\deg{\hbox{$^\circ$}}

\def\fdg{\hbox{$.\!\!^\circ$}}  
\def\farcm{\hbox{$.\mkern-4mu^\prime$}}

\def \hi {H\,{\sc i~}} 
\def \hii {H\,{\sc ii~}} 

\def\NH{$N_{\rm HI}$} 

\def\kms{km\,s$^{-1}$} 

\begin{document} 

\title{An automated search for high--velocity clouds in the
  Leiden/Dwingeloo Survey 
\thanks{Table~\ref{table:HVC} is only available in electronic form at
  the CDS via anonymous ftp to cdsarc.u-strasbg.fr (130.79.125.5) or
  via http://cdsarc.u-strasbg.fr/Abstract.html} }

\titlerunning{An automated search for HVCs in the LDS} 

\author{V. de Heij\inst{1}, R. Braun\inst{2}, and W. B. Burton\inst{1}}

\authorrunning{V. de Heij et al.}

\institute{Sterrewacht Leiden, P.\,O. Box 9513, 2300 RA Leiden, 
The Netherlands\\
\email{deheij@strw.leidenuniv.nl; burton@strw.leidenuniv.nl} 
\and Netherlands
Foundation for Research in Astronomy,    P.\,O. Box 2, 7990 AA Dwingeloo,
The Netherlands\\
\email{rbraun@nfra.nl}} 

\date{Received mmddyy/ Accepted mmddyy} 

\offprints{V. de Heij}

\abstract{ We describe an automated search through the Leiden/Dwingeloo
\hi Survey (LDS) for high--velocity clouds north of
$\delta=-28^\circ$. From the general catalog we extract a sample of
isolated high--velocity clouds, CHVCs: anomalous--velocity \hi
clouds which are {\it sharply bounded in angular extent} with no
kinematic or spatial connection to other \hi features down to a
limiting column density of 1.5$\times$10$^{18}$ cm$^{-2}$. This column
density is an order of magnitude lower than the critical \hi column
density, $\sim$2$\times$10$^{19}$ cm$^{-2}$, where the ionized fraction
is thought to increase dramatically due to the extragalactic radiation
field. As such, these objects are likely to provide their own shielding
to ionizing radiation. Their small median angular size, of about
$1^\circ$ FWHM, might then imply substantial distances, since the
partially ionized \hi skin in a power--law ionizing photon field has a
typical exponential scale--length of 1~kpc. The automated search
algorithm has been applied to the HIPASS and to the Leiden/Dwingeloo
data sets.  The results from the LDS are described here; Putman et
al. (\cite{putman02}) describe application of this algorithm to the
HIPASS material.  We identify 67 CHVCs in the LDS which satisfy
stringent requirements on isolation, and an additional 49 objects which
satisfy somewhat less stringent requirements.  Independent confirmation
is available for all of these objects, either from earlier data in the
literature or from new observations made with the Westerbork Synthesis
Radio Telescope and reported here.  The catalog includes 54 of the 65
CHVCs listed by Braun \& Burton (\cite{braun99}) on the basis of a
visual search of the LDS data.  \keywords ISM: clouds -- ISM:
kinematics and dynamics -- Galaxy: evolution -- Galaxies: dwarf --
Galaxies: evolution -- Galaxies: Local Group }

\maketitle

\section{Introduction} 
High--velocity clouds (HVCs) were first encountered in the
$\lambda\,21$ cm line of \hi at radial velocities unexplained by any
conventional model of Galactic rotation.  Since their discovery by
Muller et al.~(\cite{muller63}), they have remained enigmatic objects
of continued interest.  Wakker \& Van Woerden~(\cite{wakker97}) and
Wakker et al.~(\cite{wakker99}) have given recent reviews; since these
reviews, progress has been made on several fronts.  The
anomalous--velocity clouds are found scattered over the entire sky, and
examples are found throughout a range of radial velocity spanning about
800 \kms: obtaining an adequate observational foundation for the
phenomenon has been a persistent and continuing challenge.  We describe
here a search algorithm which has been applied to the all--sky coverage
afforded by the new \hi surveys of the northern and southern skys, and
the results of its application to the Leiden/Dwingeloo Survey for 
examples of the phenomenon.

During the past forty years, a wide variety of explanations for the
HVCs has been suggested.  The matter of distances has remained
particularly difficult, and it is on distances that most of the basic
physical properties depend. Only in a few cases have distances been
measured or constrained.  The distinct system of anomalous--velocity
features recognised as the Magellanic Stream represents tidal debris
originating in a gravitational interaction of the Large and Small
Magellanic Clouds with our Galaxy (see Putman \& Gibson
\cite{putman99}), and is therefore likely to be located at distances of
several tens of kpc.  Other distinct systems of high--velocity objects
constitute a few complexes, stretching over regions of some tens of
square degrees.  One of these, Complex A, has been found from
absorption--line observations (van Woerden et al. \cite{vanwoerden99};
Wakker \cite{wakker01}) to lie within the distance range $8 < d < 10$
kpc.  But the term HVC has been used to encompass a wide range of
phenomena; unlike the Magellanic Stream and the half--dozen well--known
complexes, many of the individual anomalous--velocity features are
compact and are isolated on the sky down to low column density limits,
as we will demonstrate below.  The properties of anomalous--velocity
\hi emission might be more readily determined after a classification
into sub--categories has been made.  After compiling a general catalog
of high--velocity features in the northern sky, we focus on identifying
the category of compact, isolated features, which show no connection in
position and velocity with the Galaxy, the Magellanic Clouds or the
extended HVC complexes.  Braun \& Burton (\cite{braun99}) have argued
that these objects may represent a single class of clouds, whose
members originated under similar circumstances and which share a common
evolutionary history, and which might lie scattered throughout the
Local Group.

The idea that the anomalous--velocity clouds are deployed throughout
the Local Group has been considered earlier, by (among others) Oort
(\cite{oort66}, \cite{oort70}, and~\cite{oort81}),
Verschuur~(\cite{verschuur75}), Eichler~(\cite{eichler76}), Einasto et
al. (\cite{einasto76}), Giovanelli (\cite{giovanelli81}),
Arp~(\cite{arp85}), and Bajaja et al. (\cite{bajaja87}).  Various
arguments have been raised against these interpretations.  In the first
review of the possible interpretations of high--velocity clouds, Oort
(\cite{oort66}) ruled out the supposition that the clouds could be
independent systems in the Local Group on two principal grounds: he
stated that `` ... a situation outside our Galaxy would give no
explanation of the principal characteristic of the high--latitude clouds,
viz. that the high velocities ... are all negative", and furthermore that
`` ... it would be almost impossible to explain on this hypothesis
high--velocity clouds which apear to be related with each other over
regions $30\deg$ or more in diameter".  Since Oort's first review, newer
\hi surveys have extended the sky coverage and have revealed that there
are, in fact, approximately as many (compact) anomalous--velocity clouds
at positive velocities as there are at negative velocities; and the
objection against the large angular size of the complexes is confronted by
the knowledge that these features, in any case, are indeed located
within the Galactic halo.

By analyzing the stability of a median HVC (in the Wakker \& Van
Woerden \cite{wakker91} tabulation) against Galactic tidal disruption
and self--gravity, Blitz et al.~(\cite{blitz99}) suggest a distance
of~1~Mpc, for an assumed ratio between \hi mass and total mass of~0.1.
Furthermore they suggest that the preferred coordinate system for the
clouds is neither the Local Standard of Rest system, nor the Galactic
Standard of Rest system, but the Local Group Standard of Rest system.
The amplitude of the average velocity and the velocity dispersion of the
cloud system both have the lowest values in this system, indicating that
it might be the most relevant.  A numerical simulation of the dynamics of
a population of low mass test masses within the gravitational potential of
the Milky Way and M~31, reproduces some aspects of the kinematic and
spatial distribution of the clouds.  They suggest that the HVCs are the
unused building blocks of the Local Group, falling towards its
barycenter.

Braun \& Burton~(\cite{braun99}, hereafter BB99)   reached  similar
conclusions based on a study of a distinct subset of the HVC
population.  By restricting their attention to compact, isolated CHVCs, they
exclude the contribution of the nearby, less representative clouds.  The
hypothesis is that the compact clouds might be the distant counterparts of
the nearby, large angular size complexes. The compact sample also shows a
natural preference for the Local Group Standard of Rest system, wherein its
velocity dispersion ($88\rm\;km\;s^{-1}$) is lower than in either the LSR or
GSR frames. The CHVCs even allow definition of a new coordinate system in
which a global minimum of the velocity dispersion ($69\rm\;km\;s^{-1}$) is
obtained. This system agrees with the Local Group system at about the
one sigma level.  Furthermore, analysis of high resolution images of
sixteen of the CHVCs provide several independent indications of distances of
between 150 and 850 kpc (Braun \& Burton~\cite{braun00}; Burton et al.
\cite{burton01a}, \cite{burton01b}). 

The BB99 sample was obtained by visual inspection of the
Leiden/Dwingeloo Survey (LDS) of the local \hi sky carried out by
Hartmann \& Burton (\cite{hartmann97}).  The LDS surveyed the sky as
far south as the Dwingeloo horizon, that is to a declination
of~$-30^\circ$; lacking information on the more southern declinations,
the BB99 conclusions were based on an incomplete sample.  A major
improvement of the CHVC study would be an extension to the whole
sky. Its high sensitivity and fully--Nyquist sampling makes the
recently completed Parkes All--Sky Survey, HIPASS, (Barnes et~al.
\cite{barnes01}) ideal for extending the CHVC sample.  To create an
all--sky resource which is as homogeneous as possible, an automated
algorithm has been developed and is described here.   This paper also 
discusses  application of the algorithm to the LDS;  a separate paper
(Putman et~al. \cite{putman02}) gives the results from applying the algorithm
to the HIPASS southern--hemisphere data.

Our discussion is organized as follows. We begin by describing the data
used and the importance of obtaining confirming observations in
\S\ref{sect:data}, proceed with a description of the algorithm and
selection criteria in \S\ref{sect:algorithm}, present a catalog of
both compact and extended high--velocity clouds in
\S\ref{sect:results}, and conclude with a brief discussion of the
global properties of the cataloged objects in
\S\ref{sect:discussion}.

\section{Observations}\label{sect:data}
The LDS was observed with the 25--m Dwingeloo telescope, whose FWHM beam
subtends $36$ arcminutes, on a grid of $0\fdg5$ by $0\fdg5$ true--angle
separation.  It covered the sky north of declination $-30\deg$ completely on
this grid, and extended in a less complete fashion a few degrees further
south.  The effective velocity coverage of the LDS spans Local
Standard of Rest velocities from $-450$
\kms~to $+400$ \kms, resolved into spectral channels of 1.03 \kms~width. 
The nominal brightness--temperature sensitivity of the LDS is 0.07 K,
although this value varies for individual spectra.  Hartmann et
al.~(\cite{hartmann96}) describe the corrections applied to the LDS
material in order to remove contamination by stray radiation.

There can be various causes of imperfections in a survey such as the
Leiden/Dwingeloo one; of these, radio frequency interference (RFI) is
the most pernicious, as it can be responsible for false detections of
just the sort of spectral signals being sought in this analysis, namely
features of quite weak intensity, moderate--to--substantial frequency
width, and occuring in only one spectrum or in a few adjacent spectra. 
More commonly, RFI produces extremely narrow spike signals, or a
characteristic $\sin(x)/x$ ringing, and can be rather easily recognised
by these and other properties as not being of an interstellar
origin. But Hartmann~(\cite{hartmann94}) and Hartmann \& Burton
(\cite{hartmann97}) show examples of RFI signals detected in the LDS 
which mimic the properties of spectral features of astronomical
interest. In these examples, the interference was of short temporal
duration, thereby disabling the common diagnostic tool of being on the
look--out for features remaining at a constant frequency, or with an
unusual telltale drift in frequency.

Because the isolated objects being sought here appear at only several
of the LDS lattice points, or even at only one, confirmation was sought
by independent observations. This policy of demanding independent
confirmation in all cases had lead BB99 to reject many candidate CHVCs
from their listing. BB99 had been able to carry out the independent
confirmations using either the NRAO 140--foot telescope or the
Dwingeloo 25--meter; but since neither of these instruments is
currently operative, we sought confirmations in new observations, made
with the Westerbork Synthesis Radio Telescope (WSRT).  The importance
of the confirmation observations is stressed by the fact that only~116
of the 171~candidate isolated features which we found in the LDS could
be verified.  Some candidates, like the one shown in
Fig.~\ref{fig:WSRT}, were revealed to be due to RFI contamination and
thus spurious, despite the fact that their spectral properties show
similarities with those of a genuine astronomical feature.  Other
candidates masquerading as astronomical features were attributed to the
vagaries of noise.  (Knowing that confirmation would be demanded, the
original list of 171 candidates was prepared with rather liberal noise
constraints.)

\begin{figure}
\caption{Two examples illustrating the role played by independent confirming 
observations made using the Westerbork Synthesis Radio Telescope in
total--power mode.  The triplet of panels on the top shows
the LDS sky image of a candidate CHVC; the lower of the spectra shows the
signal as it appears in the LDS. The upper spectrum shows the deeper
observation of the same position as carried out with the WSRT: the LDS
candidate was not confirmed, but was evidently a consequence of radio
frequency interference.  The triplet of panels on the bottom shows that the
candidate feature, CHVC\,099+07$-$356, could be confirmed using the WSRT,
which provided the upper righthand spectrum. The WSRT spectra used for the
confirmation of all LDS candidates for which there were no other independent
data were substantially more sensitive than the LDS material. The crosses on
the lower left of the sky images show the angular extent of a true degree on
the sky.  Contours are drawn in these images at~50\% and~25\% of the peak
value of the signal perceived from the candidate cloud; the gray color--bar
indicates scaling in units of K\,\kms.}
\label{fig:WSRT}
\end{figure}

Confirming observations were deemed unnecessary for those candidate
features which could be identified with objects listed in previously
published investigations based on independent data.  Most important in this
regard are the high--velocity cloud catalogs of Wakker \& van Woerden
(\cite{wakker91}), extracted from the surveys of Hulsbosch \& Wakker
(\cite{hulsbosch88}) and Bajaja et al. (\cite{bajaja85}); the BB99
catalog of CHVCs extracted from the Leiden/Dwingeloo survey and
reconfirmed with additional observations; and the catalog of Putman et
al. (\cite{putman02}) extracted from the HIPASS data using the algorithm
described here.

The differences in angular and velocity resolutions, sampling
intervals, and sensitivities of the surveys required different criteria
to ascertain matches with the features identified as candidates by the
search algorithm.  Identification of the candidates with the BB99
objects is straightforward, because the LDS material serves as input in
both cases; no additional confirmation was required, because BB99 had
already adequately confirmed the signals.  Establishing correspondence
with objects in the Putman et al.  catalog is also straightforward.
The HIPASS material was Nyquist sampled at the angular resolution
of~$15\farcm5$ afforded by the Parkes 64--m telescope, and at a
$5\sigma$~rms brightnes--temperature sensitivity of approximately 50~mK
over 26 \kms.  In these observational parameters the HIPASS data
surpasses the LDS data, and they suffice as independent confirmation of
candidates identified in the LDS.  We note, however, that the 26
\kms~velocity resolution of the HIPASS data is substantially coarser
than the 1.03 \kms~resolution of the LDS; thus it is possible that an
object of narrow linewidth would be detected in the LDS but would be
diluted by as much as a factor of 25 in the coarser HIPASS velocity
coverage. There are, in fact, two objects listed in the CHVC catalog
given in Table~\ref{table:CHVC} which lie at declinations in the overlap zone,
$-30\deg < \delta < +2\deg$, but which are not listed in the Putman et
al. catalog.

Establishing correspondence with objects listed in the Wakker \& van
Woerden (\cite{wakker91}) catalog is less straightforward, because that
listing is based on \hi observations made at substantially coarser
angular and spectral sampling than pertain to the LDS (although at
comparable sensitivity, if measured after the surveys are convolved to
similar spectral and angular resolutions).  The Hulsbosch \& Wakker
(\cite{hulsbosch88}) data, on which most of the Wakker \& van Woerden
catalog depends, pertains to observations made using the Dwingeloo
25--meter telescope at declinations above $-18\deg$, at a velocity
resolution of 8.2 \kms, and on a $1\deg$ by $1\deg$ grid; the southern
material of Bajaja et al.  (\cite{bajaja85}) was observed on the IAR
100--foot telescope at somewhat lower rms sensitivity then the
Hulsbosch \& Wakker survey and at comparable velocity resolution, but
on a coarser sampling grid of $2\deg$ by $2\deg$.

Candidates identified by our search algorithm in the LDS were judged to
correspond with objects in the catalog of Wakker \& van Woerden if the
spatial separation of the candidate and the Wakker \& van Woerden
listing does not exceed $1\deg$ and if, in addition, the velocity
separation does not exceed 25 \kms; furthermore, the velocity
difference was required to be less than the velocity dispersion of the
Gaussian which fits the central spectrum: the clouds in the Wakker \&
van Woerden catalog were constructed from the Gaussian profile
decomposition components listed by Hulsbosch \&~Wakker.  A
correspondence between the profile components and the LDS candidate
which met these criteria was judged as independent confirmation.  Of
all of the objects cataloged in Table~\ref{table:CHVC}, i.e. including
the CHVC, :HVC, and ?HVC categories defined below, 33 of the 116 could
not be identified with a Wakker \& van Woerden cloud; 28 of the 67
objects classified as CHVCs do not have a Wakker \& van Woerden
counterpart.  In all of the cases where there is an identifiable
counterpart, the coarser resolution of the Wakker \& van Woerden
material precludes measuring the degree of compactness or isolation:
these measures in Table~\ref{table:CHVC} are based on the LDS data.

For the confirming observations which we required for all of the
candidates with no definite counterpart in earlier data, we used the
Westerbork Synthesis Radio Telescope in the newly--available total--power
observing mode whereby high--resolution auto--correlation spectra are
obtained from all 14 individual 25--meter antennas, rather than the more
usual cross-correlation spectra. The position--switching mode involved
observing {\sc on}--spectra with the antennas pointed toward the
candidate and {\sc off}--spectra pointing at the same declination but
with right ascension offsets of~$\pm 3^\circ$.  The data from all 14
different telescopes and both linear polarisations were averaged into a
single spectrum, after obvious interference signals and other forms of
unreliable data had been removed. Finally, the ({\sc on}\,--\,{\sc
off})\,/\,{\sc off} spectrum was determined, using the average of both
{\sc off} spectra.  With an average rms noise of~0.02~K over 1.03~\kms,
the WSRT spectra are substantially more sensitive than the original
Leiden/Dwingeloo spectra for which confirmation was being sought.
Figure~\ref{fig:WSRT} displays two WSRT and LDS pairs, one of which
provided confirmation and one of which revealed RFI contamination
perniciously mimicking a compact high--velocity cloud.

\section{Algorithm and selection criteria}\label{sect:algorithm}

\subsection{Algorithm}
A quantified, automated routine for extracting HVCs should be designed
such that it can be applied in a general way, i.e. to surveys other
than the LDS; in that way it can also permit analysis of sample
completeness.  There are several different options for extracting
structure from a three--dimensional data cube, each having certain
advantages and disadvantages.  By using a predefined cloud model, for
example, one could decompose the data into a set of clouds which
conform to that predefined model. The input cloud model can be
described by a parametric function which is subsequently fit to the
data.  Because the shape of each cloud is presumed known, one is able,
in the context of that presumption, to handle blended emission from two
or more clouds.  An example of this approach is given by Stutzki
\&~G\"usten (\cite{stutzki90}), who used a Gaussian parametric form to
unravel C$^{18}$O emission from molecular clouds.  Alternatively, one
could create a predefined set of various possible clouds.  Thilker
et~al. (\cite{thilker98}) designed such an algorithm, and applied it to
look for \hi bubbles blown by supernovae in external galaxies.

A different approach, not based on an a priori cloud model, was used by
Williams et~al. (\cite{williams94}), among others.  Williams et
al. defined a set of contours of constant intensity, and then scanned
their molecular--cloud data for clumpy structure. Starting at a high
intensity level, a closed contour only contains the peak of a clump;
by slowly decreasing the contour level, the exact shape of the cloud
emerges. If the emission of a nearby cloud shows up in the contour, a
friend--of--friend algorithm can be used to determine to which cloud
each pixel belongs.  To extract clouds properly, the difference between
adjacent contour levels has to be small: otherwise, two nearby clouds
which each contribute local intensity peaks but with only small
differences in the peak values will be extracted as a single structure.

Following the Williams et~al. (\cite{williams94}) approach, we also use
a procedure without a presumed cloud model. But in our approach,
instead of scanning for clouds along contours of constant intensity at
varying levels, we use the gradient of the intensity field to determine
the structure to which the pixels should be assigned. By assuming that
the pixels belong to the same structure as their brightest neighbours,
we are able to extract clouds of  arbitrary shapes and sizes.

The procedure is illustrated by Fig.~\ref{fig:pixels}.  Starting at any
pixel, we proceed to its brightest neighbour, and then keep
continuously moving to the brightest neighbour of each pixel we pass,
until a local maximum is found.  The local maximum defines the peak of
the structure of which the complete track followed is part. Applying
this procedure to all pixels enables us to split up the complete
three--dimensional data set into structures, i.e. clouds. The exact
result depends on which of the adjacent pixels are called neighbours.
One could confine the set of neighbours to the adjacent pixels with a
difference in one, two, or three coordinate values of the
three--dimensional set. During our search for isolated high--velocity
clouds, we treated all adjacent pixels as neighbours, which are 26 in
number for a three--dimensional data set.

\begin{figure}
\caption{Example in two dimensions illustrating the assignment of pixels
to local maxima and the resulting definition of clouds.  Starting at the
pixels at low intensities (but above the 1.5$\sigma$ level), the algorithm
finds the appropriate maximum by moving from brightest neighbour to
brightest neighbour.  Should a pixel have two equally bright neighbours, the
assignment is random.  The righthand panel shows the result of the pixel
assignment.}\label{fig:pixels}
\end{figure}

Each local maximum in the data is, formally, the peak of a cloud, even
if the local maximum is just a small wiggle in a spectrum produced, for
example, by the vagaries of Gaussian noise. To reduce the effect of
local maxima with a low contrast compared to their environment,
adjacent clouds with low contrast were merged: specifically, two
adjacent clouds were merged if the level of the brightest intensity
contour which enclosed the peaks of both clouds exceeded either~0.2~K
or 50\%~of the intensity of the brightest peak. The brighter of the
peaks of the merged structures was considered the peak of the new
cloud.  If a cloud was encompassed by more than one neighbour, it was
joined with the neighbour with the lowest contrast.  Once a new cloud
was formed, we checked if neighbouring structures should then be
merged, according to the criterion described, with this new cloud.

Not all pixels were assigned to a cloud. We only assigned pixels with a
signal--to--noise ratio of at least~1.5 to a cloud.  Before the clouds
were merged, we required that the intensities of their peaks should
exceed three times the noise value.  After all clouds were merged, we
required furthermore that the peaks have a signal--to--noise ratio of
at least five.  As a consequence, a cloud candidate with a
signal--to--noise ratio of three which is not merged with other clouds
was removed from the list.  To determine the correct intensities for
the pixels which are formed into clouds and the peak intensities of the
clouds before merging, we used a fixed, preset noise value valid for
the complete survey.  The signal--to--noise ratio of the peaks of the
clouds after merging was determined from a locally--measured noise
value.  This value was determined in a square measuring 31 by 31
pixels, centered on the cloud and located in the velocity channel map
containing the peak intensity of the cloud in question. A lower limit
was used for this newly determined noise value, equal to the preset
noise value.

An interative procedure was used to determine the noise in the region
around the cloud peak.  We started with a sample consisting of all
pixels in the 31--by--31 pixel square. 
After determining the median, $\mu$, and the absolute deviation,
$\delta$, of the pixels considered, all pixels which were not in the
range $\mu\pm f\cdot\delta$ were removed from the sample.  By repeatedly
applying the rejection criterion, we created a sample for which all
pixels lay in the range $\mu\pm f\cdot\delta$.  The noise was then set
equal to the standard deviation of this sample.  The exact result depends
heavily on the chosen value of the factor~$f$.  The more real emission
there is in the sample, the lower the value of~$f$ should be.  We have
used the value $f=3.0$, which was found to give reasonable results.

\subsection{Application of the search algorithm to the LDS
\label{subsec:select}}

The algorithm described above can be applied generally, i.e. to a
variety of data sets.  We describe here its application to the
Leiden/Dwingeloo \hi survey.  The LDS was prepared for the algorithm as
follows.  It was first divided into 24~separate data cubes, each cube
spanning an area of extent $128^\circ$ by~$128^\circ$, and representing
the \hi sky in a zenith equal--area projection.  The sky area beyond
the inner $64^\circ$ by $64^\circ$ overlaps with neighbouring cubes.
We constructed separate cubes for the positive and for the negative
Local Standard of Rest velocities.

For the initial pass of the algorithm, the data cubes were Hanning
smoothed, with the twofold motivation of reducing the detections of
apparent clouds contributed by local maxima in the noise fluctuations
and in order to reduce the amount of   computer memory required to a
level consistent with our capabilities.  The data were smoothed with a
Gaussian function with a FWHM of 12 \kms~in the spectral direction and
$1\fdg2$ along the spatial axes.  A constant rms noise value of~0.01~K
was used for the preset noise parameter. Once all pixels with a
sufficient brightness temperature were assigned to clouds, the Hanning
smoothed data (angular sampling $0\fdg5$, velocity resolution 2.06
\kms) were used to derive the cloud properties.  

A velocity--integrated intensity map of each cloud was also extracted
from the data.  The range of integration extends over the velocity range
of the pixels that were assigned to a particular cloud. A description of all
derived cloud parameters is listed in \S\ref{sect:results}.  After lists
of clouds and their properties were produced for all of the separate 
cubes, they were merged into one catalog.  During the merging process
double entries which were found in the regions of overlapping cubes were
eliminated.

Initial application of the search algorithm to the LDS resulted in a
list of all objects satisfying the search criteria, and thus included
not only compact high--velocity clouds, but also structures that are
part of the high--velocity--cloud complexes, the intermediate--velocity
features, and even the gaseous disk of our Galaxy, as well as
features which were subsequently eliminated as due to excessive noise,
radio interference, the non--square response of the receiver bandpass, or
other imperfections in the data.  To remove emission associated with our
Galaxy and the intermediate--velocity complexes, all clouds with a deviation
velocity~$V_{\rm DEV}$ less than 70 \kms~were removed. The deviation
velocity, as defined by Wakker (\cite{wakker90}), is the excess velocity of
a feature compared to the velocities allowed by a simple model of the
kinematics of our Galaxy. A description of the model used here to define the
deviation velocity is given in the following subsection.

In order to remove the putative clouds associated with imperfections in
the data, an additional signal--to--noise criterion was adopted, namely
that the line integral of the cloud in the spectrum which passes
through the cloud peak should exceed the $8\sigma$ value.  To determine
the noise in the spectrum, all channels identified with the feature and
with a deviation velocity less than 100 \kms~were excluded from
consideration.  The iterative procedure described in the previous
section was then followed to determine $\sigma$. Features with $\int
I{\rm d}V/(\sqrt n_{\rm V} \cdot \sigma \cdot \Delta V) < 8$, (where
$n_{\rm V}$ is the number of summed velocity channels of width $\Delta
V$) were removed from the list.

The low--frequency edge of the receiver bandpass used in the LDS had a
strong roll-off beyond about $+400$ \kms, which resulted in the
occasional appearance of spectral features at the positive--velocity
extremes which could be confused with one wing of a cloud of
astronomical interest. The situation at the extreme positive velocities
in the LDS is further confused by the relatively frequent occurance of
interference in this regime.  Although the positive--velocity wings of
the LDS spectra published by Hartmann \& Burton~(\cite{hartmann97}) had
already been truncated at a~$V_{\rm LSR}$ of $+400$ \kms, in
recognition of the bandpass and RFI problems (even though the nominal
response of the receiver extended to $+500$ \kms), it seemed practical
to adopt an even lower upper--velocity limit for this project.
Therefore all objects with~$V_{\rm LSR}$ greater than $+350$ \kms~were
excluded from the list.  This selection is, however, probably without any
consequence for the high--velocity--cloud phenomenon.  The most extreme
positive velocity of the CHVC ensemble found by BB99 in the northern
hemisphere was~$+216$ \kms; the most extreme positive velocity found in
this analysis is $+268$, namely for :HVC\,$357.5+05.6+268$.  The southern
hemisphere CHVC listing is particularly relevant in this regard; a
deployment of objects scattered throughout the Local Group and with a
net infall motion (see BB99) would result in more objects at positive
velocities in the southern hemisphere than in the northern.
Nevertheless, although relatively more positive--velocity CHVCs were indeed
found by the Putman et al.  (\cite{putman02}) search through the HIPASS
material (using the algorithm described here) over the range $-700<V_{\rm
LSR}<+500$ \kms, only one of the objects identified as a CHVC had a
positive velocity more extreme than +350
\kms, namely CHVC\,$258.2-23.9+359$.  

The LDS sampled the sky on the complete $0\fdg5 \times 0\fdg5$ grid
down to a declination of~$-30^\circ$, with some observations on an
incomplete grid extending several degrees further south. The degree of
isolation of an object can only be determined if the surroundings are
well observed.  Because the information on the surroundings could not
be determined close to the edges of the survey coverage, no new
anomalous--velocity object with a declination less than~$-28^\circ$ was
entered in the catalog as a CHVC.  (Four entries in
Table~\ref{table:CHVC} have $\delta<-28\deg$: numbers 107 and 109
correspond to two objects discussed by BB99 --- their numbers 56 and
58, respectively --- and these had been subject by BB99 to new
Dwingeloo observations on a Nyquist grid, confirming their
classification; number 113 is confirmed by other data as indicated in
Table~\ref{table:CHVC}; and number 114 lies in a region where the LDS
is complete, despite the low declination.)

The degree of isolation of the clouds was determined from the
velocity--integrated images covering, an area measuring
$10\deg$ by $10\deg$ centered on the position and velocity of the cloud
in question.  The velocity interval of each image was matched to the
entire velocity extent of the object in question.  An ellipse was fit
to the contour with a value of half the maximum brightness of the cloud
to allow tabulation of size and orientation. The degree of isolation is
assessed on the basis of the lowest significant contour level of \hi
column density commensurate with the data sensitivity. Given the median
FWHM linewidth of about 25~\kms, this corresponds to a 3$\sigma$ level
of about 1.5$\times$10$^{18}$ cm$^{-2}$. Although there is a small
variation of the 3$\sigma$ \NH\ level with object linewidth, we chose
to keep this value fixed for the purposes of uniformity.  We demanded
that this contour satisfy the following criteria: (1) that it be
closed, with its greatest radial extent less than the $10\deg$ by
$10\deg$ image size; and (2) that it not be elongated in the direction
of any nearby extended emission.

A slightly different criterion for isolation was employed by Putman et
al. \cite{putman02} in their analysis of the HIPASS sample of
HVCs. Rather than employing a fixed minimum column density contour to
make this assessment, they employed the contour at 25\% of the peak
\NH\ for each object. Since the majority of detected objects are
relatively faint, with a peak column density near 12$\sigma$, the two
criteria are nearly identical for most objects. Only for the brightest
$\sim$10\% of sources might the resulting classifications differ. 

Figure~\ref{fig:designation} shows an example for each of the groups of
the classification.  Since a degree of subjectivity is involved in the
assessment of elongation with respect to nearby emission, two of the
authors (VdH and RB) each independently carried out a complete
classification of the 1280 cloud candidates. Identical independent
classifications were made in 89\% of all cases. A consensus was reached
for the remaining sources after re-examination.

\begin{figure}
\caption{Representative examples of the designation of confirmed
anomalous--velocity features into the classes of CHVC, :HVC, ?HVC, and
HVC. The crosses on the lower left of each image show the angular
extent of a true degree on the sky. The contours correspond to \NH\ =
1.5, 3, 4.5 and 6$\times$10$^{18}$ cm$^{-2}$; the gray-scale--bar
indicates scaling in units of K\,\kms.  Anomalous--velocity \hi
features designated as CHVCs are tightly constrained in their degree of
isolation at \NH=1.5$\times$10$^{18}$ cm$^{-2}$; :HVCs have some
elongation with respect to their environment; ?HVCs have an enhanced
background \NH; and HVCs are organized into large complexes at a 
significantly higher \NH=5--20$\times$10$^{18}$ cm$^{-2}$. }
\label{fig:designation}
\end{figure}

\subsection{Deviation velocity}
Every direction on the sky contains some \hi emission from the Milky Way,
which needs to be avoided when searching for high--velocity clouds.  The
kinematic and spatial properties of the Milky Way are well--enough behaved,
and well--enough known, that contamination of the anomalous--velocity sky by
\hi emission from the conventional Galactic disk can be largely avoided. 
Wakker (\cite{wakker90}) introduced the measure of deviation velocity,
defined as the difference between the velocity of the cloud and the
nearest limit of the velocities allowed by a conventional model of the
differential rotation of the Galaxy in the same direction.  Wakker's
definition of deviation velocity was based on the kinematic limits
predicted for a differentially rotating flat disk of uniform thickness.
In order to constrain possible Milky Way contamination more accurately,
we modified the definition of deviation velocity to account for the
fact that the Milky Way \hi disk is warped, and also flares,
i.e. increases in thickness with increasing Galactocentric distance,
and that it is not circular when viewed from above, but lopsided.

In order to determine the~$V_{\rm LSR}$ which corresponds to the
given~$V_{\rm DEV}$, we modeled the kinematics and spatial properties
of the Galactic \hi and calculated synthetic \hi spectra.  The range of
acceptable Galactic velocities was then set by the velocities for which
the brightness temperatures of the synthetic spectra, corresponding to
this model, exceed 0.5~K. The model consists of a thin disk which has
constant properties (central density, $n_{\rm HI}$ = 0.35 cm$^{-3}$, and
kinetic temperature, T$_k = 100$~K)
within 11.5~kpc from the Galactic center and which warps and flares
beyond that radius. The vertical $z$--distribution of the gas layer is
given by a Gaussian, with a dispersion of 180~pc for $R \le
11.5\rm\;kpc$ and increasing by 80~pc for each kpc further outward than
11.5~kpc.  This thickness is higher than that derived by Baker \&
Burton~(\cite{baker75}), for example, in order account for the fact
that the low--level wings of Galactic \hi are generally broader than
the Gaussian form exhibited at higher intensities, and thus to include
more of the low--level disk gas, especially in the outer Galaxy, into
the model.  For $R \le 11.5\rm\;kpc$ the Gaussian is centered around $z
= 0$ kpc; for $R > 11.5\rm\;kpc$, the center of the gas layer is at the
height
\begin{equation}
  z = \frac{R - 11.5}{6} \sin(\phi) + 0.3 
    \left( \frac{R - 11.5}{6} \right)^2 (1 - 2 \cos(\phi))
\end{equation}
as determined by Binney \& Merrifield (\cite{binney98}) from the Voskes
\& Burton (\cite{voskes99}) analysis of combined southern and northern
Milky Way \hi survey data. The variable $\phi$ is the galactocentric
cylindrical coordinate, which increases in the direction of Galactic
rotation and equals $180^\circ$ towards the Sun. We included an
exponential, radial decrease in density, with scale length 3.0 kpc, for
$R > 11.5$ kpc, in order to improve the resemblance between the model
and data.  The Sun is located at~8.5~kpc from the Galactic center; the
gas follows circular rotation with a flat rotation curve at the level
220 \kms. Furthermore, the synthetic \hi spectra were broadened with a
Gaussian distribution with a dispersion of 20 \kms, in order to allow,
conservatively, for the somewhat ragged edge of the outer Milky Way,
observed at low latitudes, and for the filamentary
intermediate--velocity clouds, commonly observed at higher latitudes.

Figure~\ref{fig:vdev} shows the limiting~$V_{\rm LSR}$ corresponding to
a $V_{\rm DEV}$ of~$70\rm\;km\;s^{-1}$ for this model, as function of
Galactic longitude and latitude.  The deviation--velocity approach was
adopted in order to eliminate contamination by the conventional
Galactic disk, although most models of the anomalous--velocity clouds
allow these objects to pervade velocities near zero \kms.  The
deviation--velocity approach prejudices against detecting CHVCs at low
velocities, but we note that they would likely remain unrecognized at
low velocites in any case, because of the dominance of the ubiquitous
Milky Way \hi emission.

\begin{figure}
\caption{ Velocities excluded from the search procedure, because of
possible contamination by \hi emission from the Milky Way gaseous
layer. The deviation velocity was defined to be 70 \kms~more extreme
than the empirically determined extrema of the conventional gaseous
layer, albeit warped and lopsided (Voskes \& Burton \cite{voskes99}).
The $V_{\rm LSR}$ which corresponds with a $V_{\rm DEV}$~of 70 \kms~is
shown as function of the Galactic longitude and latitude. For each line
of sight there is a negative (top) and a positive~$V_{\rm LSR}$
(bottom) corresponding with the given~$V_{\rm DEV}$.  Black contours are
drawn at~$\pm 90$ and~$\pm 100\rm\;km\;s^{-1}$; white ones, at~$\pm
150$ and~$\pm 200\rm\;km\;s^{-1}$.}\label{fig:vdev}
\end{figure}


\section{Results}\label{sect:results} 

\subsection{Catalog of high--velocity clouds}\label{subsect:table}
Table~\ref{table:HVC} lists all 917 HVCs found in the LDS with our
algorithm.  Due to the extent of the table it is only available
electronically and not in the printed version of this paper. Only the
first 15 columns (as defined below) are included for this table. The
reader should be aware that not all entries have been independently
confirmed (see the Column 15 description below). The subset of 116
objects which at least partially fulfill the criteria for source
isolation and have been confirmed in independent observations are
listed in Table~\ref{table:CHVC}. In addition to well-defined CHVCs,
other features, likewise independently confirmed, but satisfying less
stringent criteria of apparent isolation are also included in the
table.  Features designated with :HVC had some ambiguity in their
degree of isolation as ascertained with the 1.5$\times$10$^{18}$
cm$^{-2}$ \NH~contour in the LDS data. Features designated with ?HVC
had one of a number of other shortcomings as detailed below; these
include not satisfying the $V_{\rm DEV} > 70$ \kms~criterion, lying
near the edge of the LDS survey coverage either spatially or in
velocity and in some cases the presence of a significant background
level.

The columns of the table denote the following:
\begin{description}
\item[] \emph{Column~1:} Running identifying number in the catalog.
\item[] \emph{Column~2:} Designation, consisiting of a prefix, followed
by the Galactic longitude, Galactic latitude, and Local Standard of
Rest velocity.  The prefix is CHVC for the clouds satisfying both of
our isolation criteria; the prefix is :HVC for clouds for which the
isolation is less clear; the prefix is ?HVC for clouds which have some
other shortcoming (see notes) in their isolation designation; while HVC
is used to indicate clouds connected to extended complexes. The
longitude, latitude, and velocity refer to the intensity--weighted
averages of all pixels which are assigned to the cloud.
\item[] \emph{Columns~3 and~4:} J2000 right ascension and declination
      coordinates, respectively, of the position listed in column~2.
\item[] \emph{Columns~5, 6, and~7:} Radial velocities measured with
      respect to the Local Standard of Rest, the Galactic Standard of
      Rest, and the Local Group Standard of Rest systems,
      respectively. The Galactic Standard of Rest reference frame is
      defined by $V_{\rm GSR}=V_{\rm LSR}+220\cos(b)\sin(l)$; the Local
      Group Standard of Rest frame, by $V_{\rm LGSR}=V_{\rm
      GSR}-62\cos(b)\cos(l)+40\cos(b)\sin(l)-35\sin(b)$.  The input
      values of $l$, $b$, and $V_{\rm LSR}$ are those listed in
      column~2.
\item[] \emph{Column~8:} Velocity FWHM of the spectrum which passes
      through the peak--intensity pixel of the cloud.
\item[] \emph{Columns~9, 10, and~11:} Angular FWHM major axis, minor
      axis, and major axis position angle, respectively.
      Using the appropriate velocity--integrated moment--map image, an
      ellipse was fit to the contour with half the value of the maximum
      column density of the cloud (column~13).  The position angle is
      positive in the direction of increasing Galactic longitude, with
      a value of zero when the cloud is aligned pointing toward the
      Galactic north pole.
\item[] \emph{Column~12:} Peak brightness temperature of the cloud.
\item[] \emph{Column~13:} Maximum column density of the cloud in
      units of~$10^{20}\rm\;cm^{-2}$.
\item[] \emph{Column~14:} Total flux of the cloud, in units 
      of $\rm Jy\;km\;s^{-1}$.
\item[] \emph{Column~15:} Indications of occurances of the features in other
       catalogs of anomalous--velocity clouds, coded as follows, with
       the number in the appropriate catalog indicated in the table:
       {\tt WvW}, for the Wakker \& van Woerden (\cite{wakker91})
       analysis of the Hulsbosch \& Wakker (\cite{hulsbosch88}) and
       Bajaja (\cite{bajaja85}) surveys; {\tt BB}, for the BB99 visual
       search of the LDS; {\tt HP}, for the Putman et
       al. (\cite{putman02}) application of the algorithm described
       here to the HIPASS data; and {\tt WSRT}, for confirming
observations in this study.
\item[] \emph {Column 16:} References to earlier studies of individual
features, and some explanatory notes.  The references to earlier
studies of the tabulated features are coded as follows: {\tt BB00},
Braun \& Burton \cite{braun00}; {\tt BKP01}, Br\"uns et
al. \cite{bruns01}; {\tt BBC01}, Burton et al. \cite{burton01a}; {\tt
CM79}, Cohen \& Mirabel \cite{cohen79}; {\tt D75}, Davies
\cite{davies75}; {\tt G81}, Giovanelli \cite{giovanelli81}; {\tt
HSP01}, Hoffmann et al. \cite{hoffman01}; {\tt H78}, Hulsbosch
\cite{hulsbosch78}; {\tt H92}, Henning \cite{henning92}; {\tt M81},
Mirabel \cite{mirabel81}; {\tt MC79}, Mirabel \& Cohen
\cite{mirabel79}; {\tt SS74}, Saraber \& Shane \cite{saraber74}; and
{\tt W79}, Wright \cite{wright79}.  The notes carry the following
meanings: (1) denotes objects with $V_{\rm DEV}<70$ \kms~but which are
nevertheless entered in the catalog, for reasons explained in the
following subsection; (2) denotes objects at $\delta < -28\deg$ but
which are nevertheless entered in the catalog, as explained in the
following subsection; (3) denotes the object ?HVC\,$110.6-07.0-466$
which appears incompletely on the low--velocity wing of the LDS profile
but which is amply known from earlier work and therefore is entered in
the catalog; (4) denotes an object with a $T_{\rm peak}$ which falls
just below the $5\sigma$ level of the LDS but which has been amply
confirmed in the WSRT imaging of Braun \& Burton (\cite{braun00}); and
(5) denotes an object which has an enhanced background level in excess
of the nominal 1.5$\times$10$^{18}$ cm$^{-2}$ \NH\ cutoff, but for
which the background is sufficiently uniform to suggest that it might
be unrelated to the source in question.

\end{description}

\begin{table*}
\caption{Catalog of High--Velocity Clouds identified in the
Leiden/Dwingeloo Survey.}
\label{table:HVC}
\end{table*}

\begin{table*}
\caption{Isolated, high--velocity clouds identified in the Leiden/Dwingeloo Survey.}
\label{table:CHVC}

\setlength{\tabcolsep}{1mm}
\renewcommand{\arraystretch}{0.75}

\begin{tabular}{|rr|rrrrrcrrrccr|ll|}

\hline
\multicolumn{1}{|c}{                                           \tf \#} &
\multicolumn{1}{c|}{                                      designation} &
\multicolumn{1}{c}{                                            \tf RA} &
\multicolumn{1}{c}{                                           \tf DEC} &
\multicolumn{1}{c}{                        $\scriptstyle V_{\rm LSR}$} &
\multicolumn{1}{c}{                        $\scriptstyle V_{\rm GSR}$} &
\multicolumn{1}{c}{                      $\scriptstyle  V_{\rm LGSR}$} &
\multicolumn{1}{c}{                                          \tf FWHM} &
\multicolumn{1}{c}{                                           \tf MAJ} &
\multicolumn{1}{c}{                                           \tf MIN} &
\multicolumn{1}{c}{                                            \tf PA} &
\multicolumn{1}{c}{    $\scriptstyle T_{\scriptscriptstyle \rm peak}$} &
\multicolumn{1}{c}{      $\scriptstyle N_{\scriptscriptstyle \rm HI}$} &
\multicolumn{1}{c}{                                       \tf    FLUX} &
\multicolumn{1}{|l}{                                          catalog} &
\multicolumn{1}{l|}{                                       references} \\
\multicolumn{1}{|c}{                                               \ } &
\multicolumn{1}{c|}{             $\scriptstyle lll.l\pm bb.b \pm VVV$} &
\multicolumn{1}{c}{                $\scriptstyle  {\rm h}\ \ {\rm m}$} &
\multicolumn{1}{c}{                  $\scriptstyle  \circ \ \ \prime$} &
\multicolumn{1}{c}{                           $\scriptstyle \rm km/s$} &
\multicolumn{1}{c}{                           $\scriptstyle \rm km/s$} &
\multicolumn{1}{c}{                           $\scriptstyle \rm km/s$} &
\multicolumn{1}{c}{                           $\scriptstyle \rm km/s$} &
\multicolumn{1}{c}{                             $\scriptstyle ^\circ$} &
\multicolumn{1}{c}{                             $\scriptstyle ^\circ$} &
\multicolumn{1}{c}{                             $\scriptstyle ^\circ$} &
\multicolumn{1}{c}{                              $\scriptstyle \rm K$} &
\multicolumn{1}{c}{                 $\scriptstyle 10^{20}\rm cm^{-2}$} &
\multicolumn{1}{c}{                       $\scriptstyle \rm Jy\,km/s$} &
\multicolumn{1}{|l}{                                         numbers } &
\multicolumn{1}{l|}{                                       and notes } \\
\hline\hline
{\tf   1}&{\tf CHVC\,002.1\omp 03.3\omm 199}&{\tf 17 38.1}&{\tf \omm 25 26}&{\tf \omm 199}&{\tf \omm 191}&{\tf \omm 253}&{\tf  38}&{\tf 0.8}&{\tf 0.8}&{\tf       0}&{\tf 0.31}&{\tf 0.15}&{\tf  112~~}&{\ttf HP19} &   \\
{\tf   2}&{\tf CHVC\,004.4\omp 05.7\omp 202}&{\tf 17 34.5}&{\tf \omm 22 15}&{\tf      202}&{\tf      219}&{\tf      157}&{\tf  32}&{\tf 0.9}&{\tf 0.6}&{\tf     170}&{\tf 0.28}&{\tf 0.16}&{\tf  121~~}&{\ttf HP45} &   \\
{\tf   3}&{\tf CHVC\,008.1\omm 05.4\omm 217}&{\tf 18 24.4}&{\tf \omm 24 34}&{\tf \omm 217}&{\tf \omm 186}&{\tf \omm 238}&{\tf  20}&{\tf 0.8}&{\tf 0.8}&{\tf       0}&{\tf 1.26}&{\tf 0.51}&{\tf 1139~~}&{\ttf WvW307,\,HP90} & {\ttf SS74} \\
{\tf   4}&{\tf CHVC\,008.7\omm 03.8\omm 215}&{\tf 18 19.5}&{\tf \omm 23 15}&{\tf \omm 215}&{\tf \omm 182}&{\tf \omm 235}&{\tf  37}&{\tf 2.6}&{\tf 0.7}&{\tf     180}&{\tf 0.61}&{\tf 0.38}&{\tf  736~~}&{\ttf HP92} & {\ttf SS74} \\
{\tf   5}&{\tf CHVC\,014.8\omm 05.3\omm 171}&{\tf 18 37.3}&{\tf \omm 18 34}&{\tf \omm 171}&{\tf \omm 115}&{\tf \omm 161}&{\tf  58}&{\tf 0.4}&{\tf 0.4}&{\tf       0}&{\tf 0.21}&{\tf 0.11}&{\tf   41~~}&{\ttf HP153} &   \\
{\tf   6}&{\tf CHVC\,016.7\omm 25.0\omm 230}&{\tf 19 57.9}&{\tf \omm 24 45}&{\tf \omm 230}&{\tf \omm 173}&{\tf \omm 202}&{\tf  14}&{\tf 0.8}&{\tf 0.8}&{\tf       0}&{\tf 0.47}&{\tf 0.15}&{\tf   88~~}&{\ttf BB1} & {\ttf M81} \\
{\tf   7}&{\tf ?HVC\,018.3\omp 47.1\omm 147}&{\tf 15 39.7}&{\tf \omp 10 23}&{\tf \omm 147}&{\tf \omm 100}&{\tf \omm 157}&{\tf  53}&{\tf 0.8}&{\tf 0.8}&{\tf       0}&{\tf 0.23}&{\tf 0.12}&{\tf   97~~}&{\ttf WvW57,\,BB2} & {\ttf B00; (5)}\\
{\tf   8}&{\tf CHVC\,019.2\omm 19.6\omm 263}&{\tf 19 40.0}&{\tf \omm 20 39}&{\tf \omm 263}&{\tf \omm 195}&{\tf \omm 226}&{\tf  23}&{\tf 0.4}&{\tf 0.4}&{\tf       0}&{\tf 0.24}&{\tf 0.12}&{\tf   49~~}&{\ttf HP192} &   \\
{\tf   9}&{\tf :HVC\,023.0\omm 13.5\omm 285}&{\tf 19 22.6}&{\tf \omm 14 52}&{\tf \omm 285}&{\tf \omm 202}&{\tf \omm 234}&{\tf  29}&{\tf 0.4}&{\tf 0.4}&{\ttf      0}&{\tf 0.19}&{\tf 0.07}&{\tf   35~~}&{\ttf HP221} &   \\
{\tf  10}&{\tf CHVC\,023.5\omm 19.7\omm 234}&{\tf 19 47.0}&{\tf \omm 16 57}&{\tf \omm 234}&{\tf \omm 152}&{\tf \omm 179}&{\tf  23}&{\tf 2.3}&{\tf 1.4}&{\tf     140}&{\tf 0.87}&{\tf 0.41}&{\tf 1209~~}&{\ttf WvW385,\,HP223} &   \\
{\tf  11}&{\tf CHVC\,024.3\omm 01.8\omm 290}&{\tf 18 42.2}&{\tf \omm 08 31}&{\tf \omm 290}&{\tf \omm 199}&{\tf \omm 238}&{\tf  20}&{\tf 0.8}&{\tf 0.8}&{\tf       0}&{\tf 0.43}&{\tf 0.18}&{\tf  148~~}&{\ttf WvW302,\,BB3,\,HP234} &   \\
{\tf  12}&{\tf :HVC\,026.1\omm 20.0\omm 270}&{\tf 19 52.4}&{\tf \omm 14 48}&{\tf \omm 270}&{\tf \omm 179}&{\tf \omm 203}&{\tf  14}&{\tf 0.8}&{\tf 0.8}&{\tf       0}&{\tf 0.22}&{\tf 0.07}&{\tf   36~~}&{\ttf WvW385, HP246} & \\
{\tf  13}&{\tf CHVC\,028.2\omm 04.0\omm 321}&{\tf 18 57.3}&{\tf \omm 06 01}&{\tf \omm 321}&{\tf \omm 217}&{\tf \omm 250}&{\tf  27}&{\tf 0.4}&{\tf 0.4}&{\tf       0}&{\tf 0.36}&{\tf 0.18}&{\tf  107~~}&{\ttf HP268} &   \\
{\tf  14}&{\tf ?HVC\,030.4\omm 50.7\omm 129}&{\tf 21 58.9}&{\tf \omm 22 10}&{\tf \omm 129}&{\tf \omm  59}&{\tf \omm  53}&{\tf  36}&{\tf 0.8}&{\tf 0.8}&{\tf       0}&{\tf 0.73}&{\tf 0.53}&{\tf 1484~~}&{\ttf BB4,\,HP283} &   {\ttf (5)} \\
{\tf  15}&{\tf CHVC\,031.5\omm 20.1\omm 282}&{\tf 20 01.2}&{\tf \omm 10 16}&{\tf \omm 282}&{\tf \omm 175}&{\tf \omm 193}&{\tf  38}&{\tf 0.4}&{\tf 0.4}&{\tf       0}&{\tf 0.45}&{\tf 0.32}&{\tf  215~~}&{\ttf WvW386,\,BB5,\,HP294} &   \\
{\tf  16}&{\tf CHVC\,031.5\omm 46.6\omm 178}&{\tf 21 43.2}&{\tf \omm 20 19}&{\tf \omm 178}&{\tf \omm  99}&{\tf \omm  95}&{\tf  29}&{\tf 0.8}&{\tf 0.8}&{\tf       0}&{\tf 0.27}&{\tf 0.17}&{\tf  134~~}&{\ttf HP291} &   \\
{\tf  17}&{\tf CHVC\,032.0\omm 30.9\omm 311}&{\tf 20 42.2}&{\tf \omm 14 15}&{\tf \omm 311}&{\tf \omm 211}&{\tf \omm 220}&{\tf  36}&{\tf 0.8}&{\tf 0.8}&{\tf       0}&{\tf 0.25}&{\tf 0.15}&{\tf   80~~}&{\ttf WvW443,\,BB6,\,HP299} &   \\
{\tf  18}&{\tf CHVC\,033.8\omm 38.6\omm 267}&{\tf 21 13.9}&{\tf \omm 15 52}&{\tf \omm 267}&{\tf \omm 172}&{\tf \omm 173}&{\tf  10}&{\tf 0.4}&{\tf 0.4}&{\tf       0}&{\tf 0.26}&{\tf 0.06}&{\tf   13~~}&{\ttf WvW489,\,HP326} &   \\
{\tf  19}&{\tf :HVC\,033.9\omm 39.8\omm 251}&{\tf 21 18.9}&{\tf \omm 16 18}&{\tf \omm 251}&{\tf \omm 157}&{\tf \omm 157}&{\tf  16}&{\tf 0.4}&{\tf 0.4}&{\tf       0}&{\tf 0.13}&{\tf 0.04}&{\tf   12~~}&{\ttf HP322} & \\
{\tf  20}&{\tf :HVC\,036.5\omp 09.5\omm 303}&{\tf 18 23.7}&{\tf \omp 07 29}&{\tf \omm 303}&{\tf \omm 174}&{\tf \omm 206}&{\tf  62}&{\tf 2.2}&{\tf 1.3}&{\tf \omm 80}&{\tf 0.24}&{\tf 0.10}&{\tf  134~~}&{\ttf WvW254} &   \\
{\tf  21}&{\tf :HVC\,037.3\omm 14.2\omm 202}&{\tf 19 49.8}&{\tf \omm 02 40}&{\tf \omm 202}&{\tf \omm  73}&{\tf \omm  88}&{\tf  38}&{\tf 0.9}&{\tf 0.6}&{\tf      10}&{\tf 0.23}&{\tf 0.09}&{\tf   62~~}&{\ttf WvW345,\,HP345} &   \\
{\tf  22}&{\tf :HVC\,038.3\omm 10.8\omm 288}&{\tf 19 39.6}&{\tf \omm 00 16}&{\tf \omm 288}&{\tf \omm 155}&{\tf \omm 171}&{\tf  39}&{\tf 0.8}&{\tf 0.8}&{\tf       0}&{\tf 0.22}&{\tf 0.10}&{\tf   76~~}&{\ttf WvW345, HP356} & \\
{\tf  23}&{\tf CHVC\,038.5\omp 07.3\omm 314}&{\tf 18 35.4}&{\tf \omp 08 12}&{\tf \omm 314}&{\tf \omm 178}&{\tf \omm 206}&{\tf  41}&{\tf 0.8}&{\tf 0.8}&{\tf       0}&{\tf 0.20}&{\tf 0.11}&{\tf   70~~}&{\ttf WvW268} &   \\
{\tf  24}&{\tf CHVC\,038.8\omm 33.7\omm 258}&{\tf 21 02.4}&{\tf \omm 10 11}&{\tf \omm 258}&{\tf \omm 144}&{\tf \omm 144}&{\tf  29}&{\tf 0.8}&{\tf 0.8}&{\tf       0}&{\tf 0.49}&{\tf 0.25}&{\tf  210~~}&{\ttf WvW460,\,BB8,\,HP361} &   \\
{\tf  25}&{\tf CHVC\,038.9\omm 13.7\omm 233}&{\tf 19 50.9}&{\tf \omm 01 01}&{\tf \omm 233}&{\tf \omm  99}&{\tf \omm 113}&{\tf  14}&{\tf 0.8}&{\tf 0.8}&{\tf       0}&{\tf 0.79}&{\tf 0.24}&{\tf  235~~}&{\ttf WvW345,\,HP365} &   \\
{\tf  26}&{\tf CHVC\,039.0\omm 37.1\omm 239}&{\tf 21 15.1}&{\tf \omm 11 32}&{\tf \omm 239}&{\tf \omm 129}&{\tf \omm 126}&{\tf  25}&{\tf 0.8}&{\tf 0.8}&{\tf       0}&{\tf 0.34}&{\tf 0.15}&{\tf  128~~}&{\ttf WvW482,\,BB7,\,HP360} &   \\
{\tf  27}&{\tf CHVC\,039.6\omm 31.0\omm 272}&{\tf 20 53.9}&{\tf \omm 08 23}&{\tf \omm 272}&{\tf \omm 151}&{\tf \omm 152}&{\tf  25}&{\tf 1.8}&{\tf 1.5}&{\tf \omm 80}&{\tf 0.37}&{\tf 0.21}&{\tf  326~~}&{\ttf WvW442,\,BB9,\,HP370} &   \\
{\tf  28}&{\tf CHVC\,040.0\omp 07.6\omm 314}&{\tf 18 37.1}&{\tf \omp 09 44}&{\tf \omm 314}&{\tf \omm 173}&{\tf \omm 200}&{\tf  19}&{\tf 0.8}&{\tf 0.8}&{\tf       0}&{\tf 0.20}&{\tf 0.06}&{\tf   23~~}&{\ttf WvW268} & {\ttf HW88} \\
{\tf  29}&{\tf CHVC\,040.2\omp 00.5\omm 279}&{\tf 19 03.0}&{\tf \omp 06 44}&{\tf \omm 279}&{\tf \omm 137}&{\tf \omm 159}&{\tf  42}&{\tf 0.8}&{\tf 0.8}&{\tf       0}&{\tf 0.19}&{\tf 0.11}&{\tf   61~~}&{\ttf WvW289,\,BB10} &   \\
{\tf  30}&{\tf :HVC\,040.4\omm 73.4\omm 169}&{\tf 23 39.1}&{\tf \omm 23 49}&{\tf \omm 169}&{\tf \omm 128}&{\tf \omm 101}&{\tf  20}&{\tf 0.8}&{\tf 0.8}&{\tf       0}&{\tf 0.14}&{\tf 0.06}&{\tf   46~~}&{\ttf HP367} &   \\
{\tf  31}&{\tf CHVC\,041.1\omm 27.3\omm 239}&{\tf 20 43.1}&{\tf \omm 05 33}&{\tf \omm 239}&{\tf \omm 111}&{\tf \omm 113}&{\tf  26}&{\tf 0.8}&{\tf 0.8}&{\tf       0}&{\tf 0.25}&{\tf 0.11}&{\tf   59~~}&{\ttf WvW419,\,HP379} &   \\
{\tf  32}&{\tf CHVC\,042.9\omm 12.9\omm 265}&{\tf 19 55.6}&{\tf \omp 02 43}&{\tf \omm 265}&{\tf \omm 119}&{\tf \omm 129}&{\tf  51}&{\tf 0.8}&{\tf 0.8}&{\tf       0}&{\tf 0.32}&{\tf 0.16}&{\tf   95~~}&{\ttf WvW348,\,B11,\,HP384} & {\ttf G81} \\
{\tf  33}&{\tf CHVC\,042.9\omm 13.3\omm 315}&{\tf 19 56.8}&{\tf \omp 02 38}&{\tf \omm 315}&{\tf \omm 169}&{\tf \omm 179}&{\tf  27}&{\tf 1.2}&{\tf 1.0}&{\tf \omm 60}&{\tf 0.64}&{\tf 0.40}&{\tf  396~~}&{\ttf WvW348,\,B11,\,HP384} & {\ttf G81} \\
{\tf  34}&{\tf :HVC\,043.0\omm 29.9\omm 217}&{\tf 20 55.3}&{\tf \omm 05 18}&{\tf \omm 217}&{\tf \omm  87}&{\tf \omm  85}&{\tf  25}&{\tf 0.8}&{\tf 0.8}&{\tf       0}&{\tf 0.34}&{\tf 0.22}&{\tf  262~~}&{\ttf WvW445,\,HP389} &   \\
{\tf  35}&{\tf :HVC\,045.5\omm 24.5\omm 228}&{\tf 20 40.9}&{\tf \omm 00 47}&{\tf \omm 228}&{\tf \omm  85}&{\tf \omm  84}&{\tf  36}&{\tf 0.8}&{\tf 0.8}&{\tf       0}&{\tf 0.26}&{\tf 0.13}&{\tf   84~~}&{\ttf WvW409,\,HP404} &   \\
{\tf  36}&{\tf CHVC\,050.0\omm 68.2\omm 193}&{\tf 23 23.3}&{\tf \omm 19 03}&{\tf \omm 193}&{\tf \omm 131}&{\tf \omm 102}&{\tf  30}&{\tf 0.8}&{\tf 0.8}&{\tf       0}&{\tf 0.24}&{\tf 0.13}&{\tf   99~~}&{\ttf BB13,\,HP421} &   \\
{\tf  37}&{\tf CHVC\,050.2\omm 27.1\omm 274}&{\tf 20 58.5}&{\tf \omp 01 34}&{\tf \omm 274}&{\tf \omm 124}&{\tf \omm 116}&{\tf  26}&{\tf 0.9}&{\tf 0.6}&{\tf       0}&{\tf 0.19}&{\tf 0.10}&{\tf   42~~}&{\ttf WvW421,\,HP420} &   \\
{\tf  38}&{\tf :HVC\,054.1\omp 01.5\omm 197}&{\tf 19 25.9}&{\tf \omp 19 29}&{\tf \omm 197}&{\tf \omm  19}&{\tf \omm  24}&{\tf  17}&{\tf 0.4}&{\tf 0.4}&{\tf       0}&{\tf 0.30}&{\tf 0.08}&{\tf   30~~}&{\ttf WvW283} &   \\
{\tf  39}&{\tf :HVC\,057.0\omp 03.7\omm 209}&{\tf 19 23.5}&{\tf \omp 22 59}&{\tf \omm 209}&{\tf \omm  25}&{\tf  \omm 28}&{\tf  24}&{\tf 3.8}&{\tf 0.9}&{\tf      70}&{\tf 0.32}&{\tf 0.11}&{\tf  132~~}&{\ttf WvW274} & \\
{\tf  40}&{\tf :HVC\,065.9\omm 09.4\omm 273}&{\tf 20 32.2}&{\tf \omp 23 48}&{\tf \omm 273}&{\tf \omm  75}&{\tf \omm  58}&{\tf  29}&{\tf 0.9}&{\tf 0.6}&{\tf     170}&{\tf 0.22}&{\tf 0.11}&{\tf   74~~}&{\ttf WvW327} &   \\
{\tf  41}&{\tf CHVC\,069.0\omp 03.8\omm 236}&{\tf 19 49.4}&{\tf \omp 33 33}&{\tf \omm 236}&{\tf \omm  31}&{\tf \omm  19}&{\tf  26}&{\tf 0.8}&{\tf 0.8}&{\tf       0}&{\tf 0.31}&{\tf 0.15}&{\tf  114~~}&{\ttf WvW273,\,BB15} & {\ttf BB00} \\
{\tf  42}&{\tf CHVC\,070.3\omp 50.9\omm 144}&{\tf 15 48.0}&{\tf \omp 43 58}&{\tf \omm 144}&{\tf \omm  13}&{\tf \omm  30}&{\tf  16}&{\tf 1.7}&{\tf 1.1}&{\tf \omm 50}&{\tf 0.31}&{\tf 0.12}&{\tf   84~~}&{\ttf WvW44,\,BB16} &   \\
{\tf  43}&{\tf CHVC\,070.6\omp 43.8\omm 142}&{\tf 16 27.4}&{\tf \omp 45 07}&{\tf \omm 142}&{\tf        8}&{\tf \omm   4}&{\tf  22}&{\tf 0.8}&{\tf 0.8}&{\tf       0}&{\tf 0.19}&{\tf 0.06}&{\tf   30~~}&{\ttf WvW72} &   \\
{\tf  44}&{\tf CHVC\,072.0\omm 21.9\omm 333}&{\tf 21 30.2}&{\tf \omp 20 33}&{\tf \omm 333}&{\tf \omm 139}&{\tf \omm 108}&{\tf  24}&{\tf 2.0}&{\tf 1.2}&{\tf \omm 60}&{\tf 0.75}&{\tf 0.40}&{\tf  674~~}&{\ttf WvW394} &   \\
{\tf  45}&{\tf :HVC\,073.4\omp 33.3\omm 206}&{\tf 17 28.2}&{\tf \omp 47 11}&{\tf \omm 206}&{\tf \omm  30}&{\tf \omm  32}&{\tf  26}&{\tf 1.2}&{\tf 1.2}&{\tf       0}&{\tf 0.25}&{\tf 0.07}&{\tf   67~~}&{\ttf WvW90} & \\
{\tf  46}&{\tf :HVC\,076.9\omp 55.5\omm 115}&{\tf 15 16.9}&{\tf \omp 46 32}&{\tf \omm 115}&{\tf        6}&{\tf \omm   9}&{\tf  24}&{\tf 1.3}&{\tf 0.9}&{\tf      70}&{\tf 0.25}&{\tf 0.10}&{\tf  102~~}&{\ttf WvW28} &   \\
{\tf  47}&{\tf CHVC\,077.5\omm 38.9\omm 320}&{\tf 22 33.5}&{\tf \omp 11 32}&{\tf \omm 320}&{\tf \omm 152}&{\tf \omm 110}&{\tf  20}&{\tf 1.3}&{\tf 0.8}&{\tf      40}&{\tf 0.25}&{\tf 0.10}&{\tf   52~~}&{\ttf WvW485} &   \\
{\tf  48}&{\tf CHVC\,078.1\omp 44.1\omm 149}&{\tf 16 21.8}&{\tf \omp 50 23}&{\tf \omm 149}&{\tf        5}&{\tf        0}&{\tf  29}&{\tf 0.8}&{\tf 0.8}&{\tf       0}&{\tf 0.19}&{\tf 0.10}&{\tf   60~~}&{\ttf WvW70} &   \\
{\tf  49}&{\tf :HVC\,078.4\omp 54.2\omm 158}&{\tf 15 21.8}&{\tf \omp 47 49}&{\tf \omm 158}&{\tf \omm  32}&{\tf \omm  45}&{\tf  27}&{\tf 1.5}&{\tf 1.1}&{\tf       0}&{\tf 0.19}&{\tf 0.08}&{\tf   88~~}&{\ttf WvW28} &   \\
{\tf  50}&{\tf :HVC\,080.1\omp 22.3\omm 209}&{\tf 18 41.4}&{\tf \omp 50 59}&{\tf \omm 209}&{\tf \omm   8}&{\tf        5}&{\tf  20}&{\tf 1.3}&{\tf 0.9}&{\tf      70}&{\tf 0.23}&{\tf 0.07}&{\tf   65~~}&{\ttf WvW191} &   \\
{\tf  51}&{\tf CHVC\,082.2\omp 24.6\omm 196}&{\tf 18 29.8}&{\tf \omp 53 25}&{\tf \omm 196}&{\tf        2}&{\tf       16}&{\tf  20}&{\tf 1.3}&{\tf 1.3}&{\tf \omm 90}&{\tf 0.35}&{\tf 0.13}&{\tf  300~~}&{\ttf WvW182} &   \\
{\tf  52}&{\tf :HVC\,087.2\omp 02.7\omm 296}&{\tf 20 48.7}&{\tf \omp 48 01}&{\tf \omm 296}&{\tf \omm  76}&{\tf \omm  41}&{\tf  45}&{\tf 0.8}&{\tf 0.8}&{\tf       0}&{\tf 0.24}&{\tf 0.13}&{\tf  116~~}&{\ttf WvW278,\,BB17} &   \\
{\tf  53}&{\tf :HVC\,089.4\omm 65.1\omm 315}&{\tf 23 57.3}&{\tf \omm 05 46}&{\tf \omm 315}&{\tf \omm 222}&{\tf \omm 174}&{\tf  24}&{\tf 2.0}&{\tf 1.2}&{\tf      30}&{\tf 0.18}&{\tf 0.08}&{\tf   73~~}&{\ttf WvW547,\,HP495} &   \\
{\tf  54}&{\tf :HVC\,094.4\omm 63.0\omm 321}&{\tf 00 01.3}&{\tf \omm 02 58}&{\tf \omm 321}&{\tf \omm 221}&{\tf \omm 170}&{\tf  21}&{\tf 0.8}&{\tf 0.8}&{\tf       0}&{\tf 0.36}&{\tf 0.16}&{\tf  143~~}&{\ttf WvW542,\,BB20,\,HP505} &   \\
{\tf  55}&{\tf CHVC\,099.9\omm 48.8\omm 392}&{\tf 23 50.5}&{\tf \omp 11 21}&{\tf \omm 392}&{\tf \omm 249}&{\tf \omm 190}&{\tf  36}&{\tf 0.8}&{\tf 0.8}&{\tf       0}&{\tf 0.27}&{\tf 0.15}&{\tf  115~~}&{\ttf WvW493,\,BB21} & {\ttf BBC01} \\
{\tf  56}&{\tf :HVC\,101.7\omm 41.3\omm 430}&{\tf 23 44.8}&{\tf \omp 18 48}&{\tf \omm 430}&{\tf \omm 269}&{\tf \omm 207}&{\tf  45}&{\tf 0.8}&{\tf 0.8}&{\tf       0}&{\tf 0.17}&{\tf 0.08}&{\tf   36~~}&{\ttf WvW491} &   \\
{\tf  57}&{\tf CHVC\,103.4\omm 40.1\omm 414}&{\tf 23 48.0}&{\tf \omp 20 22}&{\tf \omm 414}&{\tf \omm 250}&{\tf \omm 187}&{\tf  24}&{\tf 1.9}&{\tf 0.9}&{\tf      40}&{\tf 0.30}&{\tf 0.14}&{\tf  155~~}&{\ttf WvW491} & {\ttf HSP01} \\
{\tf  58}&{\tf :HVC\,103.8\omm 48.1\omm 167}&{\tf 23 59.5}&{\tf \omp 12 49}&{\tf \omm 167}&{\tf \omm  24}&{\tf       38}&{\tf  25}&{\tf 0.4}&{\tf 0.4}&{\tf       0}&{\tf 0.21}&{\tf 0.08}&{\tf   34~~}&{\ttf WvW518} &   \\
\hline
\end{tabular}
\end{table*}

\begin{table*}
{\bf Table 2.} (continued.)

\medskip
\setlength{\tabcolsep}{1mm}
\renewcommand{\arraystretch}{0.75}

\begin{tabular}{|rr|rrrrrcrrrccr|ll|}

\hline
\multicolumn{1}{|c}{                                           \tf \#} &
\multicolumn{1}{c|}{                                      designation} &
\multicolumn{1}{c}{                                            \tf RA} &
\multicolumn{1}{c}{                                           \tf DEC} &
\multicolumn{1}{c}{                        $\scriptstyle V_{\rm LSR}$} &
\multicolumn{1}{c}{                        $\scriptstyle V_{\rm GSR}$} &
\multicolumn{1}{c}{                      $\scriptstyle  V_{\rm LGSR}$} &
\multicolumn{1}{c}{                                          \tf FWHM} &
\multicolumn{1}{c}{                                           \tf MAJ} &
\multicolumn{1}{c}{                                           \tf MIN} &
\multicolumn{1}{c}{                                            \tf PA} &
\multicolumn{1}{c}{    $\scriptstyle T_{\scriptscriptstyle \rm peak}$} &
\multicolumn{1}{c}{      $\scriptstyle N_{\scriptscriptstyle \rm HI}$} &
\multicolumn{1}{c}{                                       \tf    FLUX} &
\multicolumn{1}{|l}{                                          catalog} &
\multicolumn{1}{l|}{                                       references} \\
\multicolumn{1}{|c}{                                               \ } &
\multicolumn{1}{c|}{             $\scriptstyle lll.l\pm bb.b \pm VVV$} &
\multicolumn{1}{c}{                $\scriptstyle  {\rm h}\ \ {\rm m}$} &
\multicolumn{1}{c}{                  $\scriptstyle  \circ \ \ \prime$} &
\multicolumn{1}{c}{                           $\scriptstyle \rm km/s$} &
\multicolumn{1}{c}{                           $\scriptstyle \rm km/s$} &
\multicolumn{1}{c}{                           $\scriptstyle \rm km/s$} &
\multicolumn{1}{c}{                           $\scriptstyle \rm km/s$} &
\multicolumn{1}{c}{                             $\scriptstyle ^\circ$} &
\multicolumn{1}{c}{                             $\scriptstyle ^\circ$} &
\multicolumn{1}{c}{                             $\scriptstyle ^\circ$} &
\multicolumn{1}{c}{                              $\scriptstyle \rm K$} &
\multicolumn{1}{c}{                 $\scriptstyle 10^{20}\rm cm^{-2}$} &
\multicolumn{1}{c}{                       $\scriptstyle \rm Jy\,km/s$} &
\multicolumn{1}{|l}{                                         numbers } &
\multicolumn{1}{l|}{                                       and notes } \\
\hline\hline
{\tf  59}&{\tf CHVC\,107.7\omm 29.7\omm 429}&{\tf 23 49.3}&{\tf \omp 31 19}&{\tf \omm 429}&{\tf \omm 247}&{\tf \omm 180}&{\tf  40}&{\tf 0.4}&{\tf 0.4}&{\tf       0}&{\tf 0.29}&{\tf 0.13}&{\tf   89~~}&{\ttf WvW437,\,BB22} & {\ttf H92} \\
{\tf  60}&{\tf CHVC\,108.3\omm 21.2\omm 402}&{\tf 23 40.2}&{\tf \omp 39 40}&{\tf \omm 402}&{\tf \omm 208}&{\tf \omm 141}&{\tf  33}&{\tf 0.8}&{\tf 0.8}&{\tf       0}&{\tf 0.22}&{\tf 0.08}&{\tf   52~~}&{\ttf BB23,\,WvW389} &   \\
{\tf  61}&{\tf ?HVC\,110.6\omm 07.0\omm 466}&{\tf 23 27.1}&{\tf \omp 53 50}&{\tf \omm 466}&{\tf \omm 262}&{\tf \omm 199}&{\tf  25}&{\tf 1.0}&{\tf 0.9}&{\tf      70}&{\tf 0.17}&{\tf 0.08}&{\tf   91~~}&{\ttf WvW318,\,BB24} & {\ttf H78,\,CM79,\,WS91; (3)} \\
{\tf  62}&{\tf CHVC\,113.7\omm 10.6\omm 442}&{\tf 23 53.0}&{\tf \omp 51 13}&{\tf \omm 442}&{\tf \omm 244}&{\tf \omm 177}&{\tf  11}&{\tf 0.9}&{\tf 0.6}&{\tf       0}&{\tf 0.77}&{\tf 0.18}&{\tf  101~~}&{\ttf WvW330,\,BB25} & {\ttf H78,\,CM79,\,WS91} \\
{\tf  63}&{\tf ?HVC\,115.4\omp 13.4\omm 260}&{\tf 22 56.9}&{\tf \omp 74 33}&{\tf \omm 260}&{\tf \omm  67}&{\tf \omm  14}&{\tf  95}&{\tf 1.3}&{\tf 0.8}&{\tf \omm 80}&{\tf 0.16}&{\tf 0.30}&{\tf  375~~}&{\ttf BB26} &  {\ttf BB00; (4); (5)}\\
{\tf  64}&{\tf CHVC\,118.2\omm 58.1\omm 373}&{\tf 00 41.5}&{\tf \omp 04 39}&{\tf \omm 373}&{\tf \omm 270}&{\tf \omm 207}&{\tf  31}&{\tf 0.8}&{\tf 0.8}&{\tf       0}&{\tf 0.71}&{\tf 0.40}&{\tf  536~~}&{\ttf WvW532,\,BB27} & {\ttf MC79,\,G81} \\
{\tf  65}&{\tf CHVC\,118.5\omm 32.6\omm 386}&{\tf 00 34.1}&{\tf \omp 30 05}&{\tf \omm 386}&{\tf \omm 223}&{\tf \omm 149}&{\tf  26}&{\tf 0.4}&{\tf 0.4}&{\tf       0}&{\tf 0.15}&{\tf 0.07}&{\tf   19~~}& {\ttf WSRT} &  \\
{\tf  66}&{\tf :HVC\,119.0\omm 73.1\omm 300}&{\tf 00 46.7}&{\tf \omm 10 14}&{\tf \omm 300}&{\tf \omm 244}&{\tf \omm 191}&{\tf  34}&{\tf 0.4}&{\tf 0.4}&{\tf       0}&{\tf 0.15}&{\tf 0.09}&{\tf   19~~}&{\ttf WvW555,\,BB28,\,HP520} &   \\
{\tf  67}&{\tf CHVC\,119.2\omm 31.1\omm 384}&{\tf 00 36.2}&{\tf \omp 31 39}&{\tf \omm 384}&{\tf \omm 220}&{\tf \omm 146}&{\tf  19}&{\tf 0.8}&{\tf 0.8}&{\tf       0}&{\tf 0.33}&{\tf 0.12}&{\tf  117~~}&{\ttf WvW444,\,BB29} & {\ttf W79} \\
{\tf  68}&{\tf CHVC\,120.2\omm 20.0\omm 441}&{\tf 00 37.4}&{\tf \omp 42 47}&{\tf \omm 441}&{\tf \omm 262}&{\tf \omm 188}&{\tf  18}&{\tf 0.4}&{\tf 0.4}&{\tf \omm 80}&{\tf 0.29}&{\tf 0.10}&{\tf   22~~}& {\ttf WSRT} & {\ttf D75} \\
{\tf  69}&{\tf CHVC\,122.9\omm 31.8\omm 325}&{\tf 00 51.4}&{\tf \omp 31 04}&{\tf \omm 325}&{\tf \omm 168}&{\tf \omm  93}&{\tf  34}&{\tf 0.4}&{\tf 0.4}&{\tf       0}&{\tf 0.20}&{\tf 0.11}&{\tf   51~~}&{\ttf WvW446,\,BB30} &   \\
{\tf  70}&{\tf CHVC\,123.7\omm 12.4\omm 214}&{\tf 00 55.8}&{\tf \omp 50 26}&{\tf \omm 214}&{\tf \omm  36}&{\tf       38}&{\tf  31}&{\tf 1.2}&{\tf 1.0}&{\tf \omm 60}&{\tf 0.28}&{\tf 0.13}&{\tf  125~~}&{\ttf WvW287} &   \\
{\tf  71}&{\tf CHVC\,125.3\omp 41.3\omm 205}&{\tf 12 22.5}&{\tf \omp 75 43}&{\tf \omm 205}&{\tf \omm  70}&{\tf \omm  42}&{\tf   7}&{\tf 0.8}&{\tf 0.8}&{\tf       0}&{\tf 1.99}&{\tf 0.31}&{\tf  252~~}&{\ttf WvW84,\,BB31} & {\ttf BB00,\,BKP01} \\
{\tf  72}&{\tf CHVC\,128.6\omp 14.7\omm 306}&{\tf 02 28.5}&{\tf \omp 76 30}&{\tf \omm 306}&{\tf \omm 140}&{\tf \omm  81}&{\tf  12}&{\tf 0.8}&{\tf 0.8}&{\tf       0}&{\tf 0.50}&{\tf 0.15}&{\tf  117~~}&{\ttf WvW231,\,BB32} &   \\
{\tf  73}&{\tf CHVC\,130.0\omm 34.2\omm 367}&{\tf 01 18.0}&{\tf \omp 28 20}&{\tf \omm 367}&{\tf \omm 228}&{\tf \omm 150}&{\tf  13}&{\tf 0.8}&{\tf 0.8}&{\tf       0}&{\tf 0.42}&{\tf 0.11}&{\tf   54~~}&{\ttf WvW466} &   \\
{\tf  74}&{\tf CHVC\,130.8\omp 60.1\omm 121}&{\tf 12 22.9}&{\tf \omp 56 33}&{\tf \omm 121}&{\tf \omm  38}&{\tf \omm  33}&{\tf  17}&{\tf 1.5}&{\tf 1.1}&{\tf     150}&{\tf 0.29}&{\tf 0.12}&{\tf  154~~}&{\ttf WvW17} &   \\
{\tf  75}&{\tf :HVC\,132.0\omm 75.8\omm 304}&{\tf 01 00.6}&{\tf \omm 13 07}&{\tf \omm 304}&{\tf \omm 264}&{\tf \omm 212}&{\tf  34}&{\tf 1.7}&{\tf 1.2}&{\tf      30}&{\tf 0.20}&{\tf 0.11}&{\tf   98~~}&{\ttf WvW557,\,BB33,\,HP530} &   \\
{\tf  76}&{\tf :HVC\,132.7\omp 25.3\omm 207}&{\tf 06 12.0}&{\tf \omp 81 01}&{\tf \omm 207}&{\tf \omm  61}&{\tf \omm  11}&{\tf  11}&{\tf 1.6}&{\tf 1.3}&{\tf \omm 90}&{\tf 0.54}&{\tf 0.09}&{\tf  199~~}&{\ttf WvW117} & \\
{\tf  77}&{\tf CHVC\,136.1\omm 23.5\omm 153}&{\tf 01 52.8}&{\tf \omp 37 49}&{\tf \omm 153}&{\tf \omm  13}&{\tf       68}&{\tf  24}&{\tf 1.9}&{\tf 1.2}&{\tf      40}&{\tf 0.44}&{\tf 0.24}&{\tf  526~~}&{\ttf WvW404} &   \\
{\tf  78}&{\tf :HVC\,141.4\omm 81.9\omm 223}&{\tf 01 02.2}&{\tf \omm 19 27}&{\tf \omm 223}&{\tf \omm 204}&{\tf \omm 159}&{\tf  25}&{\tf 0.8}&{\tf 0.8}&{\tf       0}&{\tf 0.14}&{\tf 0.06}&{\tf   26~~}&{\ttf HP534} &   \\
{\tf  79}&{\tf :HVC\,145.2\omm 77.6\omm 273}&{\tf 01 10.8}&{\tf \omm 15 35}&{\tf \omm 273}&{\tf \omm 246}&{\tf \omm 196}&{\tf  24}&{\tf 0.8}&{\tf 0.8}&{\tf       0}&{\tf 0.23}&{\tf 0.12}&{\tf   92~~}&{\ttf WvW560,\,BB34,\,HP537} &   \\
{\tf  80}&{\tf CHVC\,148.9\omm 82.5\omm 269}&{\tf 01 05.5}&{\tf \omm 20 18}&{\tf \omm 269}&{\tf \omm 254}&{\tf \omm 210}&{\tf  20}&{\tf 0.8}&{\tf 0.8}&{\tf       0}&{\tf 0.39}&{\tf 0.18}&{\tf  114~~}&{\ttf BB36,\,HP538} &   \\
{\tf  81}&{\tf :HVC\,155.5\omp 04.0\omm 155}&{\tf 04 48.0}&{\tf \omp 51 17}&{\tf \omm 155}&{\tf \omm  64}&{\tf        7}&{\tf  46}&{\tf 0.8}&{\tf 0.8}&{\tf       0}&{\tf 0.64}&{\tf 0.18}&{\tf  142~~}&{\ttf WvW247} &   \\
{\tf  82}&{\tf CHVC\,157.1\omp 02.9\omm 186}&{\tf 04 48.7}&{\tf \omp 49 22}&{\tf \omm 186}&{\tf \omm 101}&{\tf \omm  30}&{\tf  12}&{\tf 0.9}&{\tf 0.6}&{\tf     180}&{\tf 0.36}&{\tf 0.10}&{\tf   49~~}&{\ttf WvW275,\,BB38} &   \\
{\tf  83}&{\tf CHVC\,157.7\omm 39.3\omm 287}&{\tf 02 40.9}&{\tf \omp 16 04}&{\tf \omm 287}&{\tf \omm 222}&{\tf \omm 144}&{\tf  12}&{\tf 0.4}&{\tf 0.4}&{\tf       0}&{\tf 0.22}&{\tf 0.08}&{\tf   24~~}&{\ttf WvW486,\,BB39} & {\ttf BBC01; HSP01} \\
{\tf  84}&{\tf CHVC\,161.6\omp 02.7\omm 186}&{\tf 05 04.9}&{\tf \omp 45 43}&{\tf \omm 186}&{\tf \omm 117}&{\tf \omm  47}&{\tf  23}&{\tf 0.8}&{\tf 0.8}&{\tf       0}&{\tf 0.64}&{\tf 0.32}&{\tf  219~~}&{\ttf WvW277,\,BB40} &   \\
{\tf  85}&{\tf CHVC\,170.8\omm 42.3\omm 217}&{\tf 03 05.8}&{\tf \omp 07 45}&{\tf \omm 217}&{\tf \omm 191}&{\tf \omm 117}&{\tf  24}&{\tf 0.8}&{\tf 0.8}&{\tf       0}&{\tf 0.43}&{\tf 0.15}&{\tf  117~~}&{\ttf WvW490} &   \\
{\tf  86}&{\tf CHVC\,171.3\omm 53.6\omm 238}&{\tf 02 36.6}&{\tf \omm 00 55}&{\tf \omm 238}&{\tf \omm 218}&{\tf \omm 150}&{\tf  28}&{\tf 0.8}&{\tf 0.8}&{\tf       0}&{\tf 0.53}&{\tf 0.28}&{\tf  341~~}&{\ttf WvW525,\,BB41,\,HP570} & {\ttf H78} \\
{\tf  87}&{\tf CHVC\,171.7\omm 59.7\omm 234}&{\tf 02 21.2}&{\tf \omm 05 36}&{\tf \omm 234}&{\tf \omm 218}&{\tf \omm 154}&{\tf  23}&{\tf 0.8}&{\tf 0.8}&{\tf       0}&{\tf 0.23}&{\tf 0.10}&{\tf   48~~}&{\ttf WvW536,\,BB43,\,HP571} &   \\
{\tf  88}&{\tf :HVC\,172.3\omm 41.9\omm 292}&{\tf 03 10.2}&{\tf \omp 07 17}&{\tf \omm 292}&{\tf \omm 270}&{\tf \omm 197}&{\tf  42}&{\tf 0.9}&{\tf 0.6}&{\tf     160}&{\tf 0.28}&{\tf 0.11}&{\tf   69~~}&{\ttf WvW501} &   \\
{\tf  89}&{\tf :HVC\,173.4\omm 51.9\omm 230}&{\tf 02 45.3}&{\tf \omm 00 31}&{\tf \omm 230}&{\tf \omm 215}&{\tf \omm 146}&{\tf  30}&{\tf 0.8}&{\tf 0.8}&{\tf       0}&{\tf 0.25}&{\tf 0.12}&{\tf   53~~}&{\ttf WvW525,\,HP572} &   \\
{\tf  90}&{\tf :HVC\,173.7\omm 40.5\omm 203}&{\tf 03 17.2}&{\tf \omp 07 32}&{\tf \omm 203}&{\tf \omm 185}&{\tf \omm 112}&{\tf  33}&{\tf 0.4}&{\tf 0.4}&{\tf       0}&{\tf 0.22}&{\tf 0.11}&{\tf   45~~}&{\ttf WvW467} & \\
{\tf  91}&{\tf CHVC\,175.8\omm 53.0\omm 216}&{\tf 02 46.1}&{\tf \omm 02 21}&{\tf \omm 216}&{\tf \omm 207}&{\tf \omm 140}&{\tf  28}&{\tf 0.8}&{\tf 0.8}&{\tf       0}&{\tf 0.20}&{\tf 0.08}&{\tf   30~~}&{\ttf WvW525,\,HP573} &   \\
{\tf  92}&{\tf CHVC\,186.3\omp 18.8\omm 109}&{\tf 07 16.5}&{\tf \omp 31 41}&{\tf \omm 109}&{\tf \omm 132}&{\tf \omm  89}&{\tf  14}&{\tf 0.9}&{\tf 0.6}&{\tf     160}&{\tf 1.10}&{\tf 0.32}&{\tf  305~~}&{\ttf WvW215,\,BB44} & {\ttf BBC01} \\
{\tf  93}&{\tf CHVC\,190.2\omm 30.5\omm 168}&{\tf 04 22.6}&{\tf \omp 03 47}&{\tf \omm 168}&{\tf \omm 201}&{\tf \omm 137}&{\tf  30}&{\tf 1.3}&{\tf 1.3}&{\tf \omm 80}&{\tf 0.55}&{\tf 0.25}&{\tf 1140~~}&{\ttf WvW467} &   \\
{\tf  94}&{\tf ?HVC\,190.9\omp 60.4\omp 093}&{\tf 10 36.9}&{\tf \omp 34 10}&{\tf       93}&{\tf       73}&{\tf       69}&{\tf  30}&{\tf 1.2}&{\tf 1.1}&{\tf      80}&{\tf 0.38}&{\tf 0.22}&{\tf  324~~}&{\ttf BB45} & {\ttf BB00; (1); (5)} \\
{\tf  95}&{\tf ?HVC\,197.5\omm 12.0\omm 106}&{\tf 05 40.2}&{\tf \omp 07 51}&{\tf \omm 106}&{\tf \omm 171}&{\tf \omm 117}&{\tf  25}&{\tf 0.8}&{\tf 0.8}&{\tf       0}&{\tf 0.48}&{\tf 0.27}&{\tf  284~~}&{\ttf WvW343,\,BB46} & {\ttf BBC01; (5)} \\
{\tf  96}&{\tf ?HVC\,200.2\omp 29.7\omp 080}&{\tf 08 22.2}&{\tf \omp 23 20}&{\tf       75}&{\tf        9}&{\tf       30}&{\tf  29}&{\tf 0.6}&{\tf 0.6}&{\tf       0}&{\tf 0.50}&{\tf 0.28}&{\tf  118~~}&{\ttf BB47} & {\ttf (1); (5)} \\
{\tf  97}&{\tf CHVC\,200.6\omp 52.3\omp 107}&{\tf 10 00.0}&{\tf \omp 28 29}&{\tf      107}&{\tf       60}&{\tf       59}&{\tf  22}&{\tf 0.8}&{\tf 0.8}&{\tf       0}&{\tf 0.37}&{\tf 0.17}&{\tf   89~~}& {\ttf WSRT} &   \\
{\tf  98}&{\tf ?HVC\,200.7\omm 16.0\omm 098}&{\tf 05 32.2}&{\tf \omp 03 13}&{\tf \omm  98}&{\tf \omm 172}&{\tf \omm 120}&{\tf  31}&{\tf 1.3}&{\tf 0.8}&{\tf      50}&{\tf 0.58}&{\tf 0.40}&{\tf  730~~}&{\ttf WvW362,\,BB48} &  {\ttf (5)} \\
{\tf  99}&{\tf ?HVC\,202.2\omp 30.4\omp 057}&{\tf 08 27.4}&{\tf \omp 21 55}&{\tf       57}&{\tf \omm  15}&{\tf        6}&{\tf  26}&{\tf 1.9}&{\tf 1.4}&{\tf      40}&{\tf 1.12}&{\tf 0.57}&{\tf 1796~~}&{\ttf BB49} & {\ttf BBC01; (1); (5)} \\
{\tf 100}&{\tf ?HVC\,204.2\omp 29.8\omp 075}&{\tf 08 27.5}&{\tf \omp 20 09}&{\tf       61}&{\tf \omm  17}&{\tf        0}&{\tf  34}&{\tf 0.8}&{\tf 0.8}&{\tf       0}&{\tf 1.19}&{\tf 0.79}&{\tf  777~~}&{\ttf BB50} & {\ttf BB00,\,BBC01; (1); (5)} \\
{\tf 101}&{\tf CHVC\,217.9\omp 28.7\omp 145}&{\tf 08 43.1}&{\tf \omp 08 42}&{\tf      145}&{\tf       26}&{\tf       31}&{\tf   7}&{\tf 0.4}&{\tf 0.4}&{\tf       0}&{\tf 0.66}&{\tf 0.09}&{\tf   56~~}&{\ttf WvW159} &   \\
{\tf 102}&{\tf CHVC\,218.4\omm 87.9\omm 260}&{\tf 01 00.7}&{\tf \omm 27 18}&{\tf \omm 260}&{\tf \omm 264}&{\tf \omm 229}&{\tf  29}&{\tf 0.8}&{\tf 0.8}&{\tf       0}&{\tf 0.35}&{\tf 0.15}&{\tf  102~~}&{\ttf BB51,\,HP615} &   \\
{\tf 103}&{\tf ?HVC\,224.6\omp 35.9\omp 082}&{\tf 09 19.2}&{\tf \omp 06 59}&{\tf       82}&{\tf \omm  43}&{\tf \omm  51}&{\tf  36}&{\tf 1.4}&{\tf 1.2}&{\tf      80}&{\tf 0.22}&{\tf 0.16}&{\tf  307~~}&{\ttf WvW115,\,BB53} & {\ttf (1); (5)} \\
{\tf 104}&{\tf CHVC\,224.6\omm 08.0\omp 188}&{\tf 06 43.6}&{\tf \omm 13 57}&{\tf      188}&{\tf       36}&{\tf       56}&{\tf  21}&{\tf 0.8}&{\tf 0.8}&{\tf       0}&{\tf 0.32}&{\tf 0.15}&{\tf  110~~}&{\ttf WvW325,\,BB52,\,HP633} &   \\
{\tf 105}&{\tf CHVC\,225.0\omm 41.9\omp 176}&{\tf 04 28.6}&{\tf \omm 26 15}&{\tf      176}&{\tf       61}&{\tf       96}&{\tf  60}&{\tf 0.8}&{\tf 0.8}&{\tf       0}&{\tf 0.34}&{\tf 0.20}&{\tf  253~~}&{\ttf BB54,\,HP639} &   \\
{\tf 106}&{\tf ?HVC\,226.5\omm 33.5\omp 101}&{\tf 05 05.9}&{\tf \omm 25 14}&{\tf      101}&{\tf \omm  32}&{\tf \omm   1}&{\tf  27}&{\tf 0.8}&{\tf 0.8}&{\tf       0}&{\tf 0.70}&{\tf 0.40}&{\tf  507~~}&{\ttf BB55,\,HP648} &  {\ttf (5)} \\
{\tf 107}&{\tf ?HVC\,228.9\omm 74.2\omm 168}&{\tf 02 02.0}&{\tf \omm 30 22}&{\tf \omm 174}&{\tf \omm 219}&{\tf \omm 182}&{\tf  33}&{\tf 0.9}&{\tf 0.8}&{\tf \omm 60}&{\tf 0.25}&{\tf 0.16}&{\tf  136~~}&{\ttf BB56,\,HP691} & {\ttf (2); (5)} \\
{\tf 108}&{\tf CHVC\,229.5\omp 60.6\omp 151}&{\tf 10 54.5}&{\tf \omp 15 51}&{\tf      151}&{\tf       69}&{\tf       43}&{\tf  23}&{\tf 0.4}&{\tf 0.4}&{\tf       0}&{\tf 0.18}&{\tf 0.08}&{\tf   28~~}&{\ttf BB57} & {\ttf BB00,\,BBC01} \\
{\tf 109}&{\tf ?HVC\,235.3\omm 73.7\omm 150}&{\tf 02 02.7}&{\tf \omm 32 10}&{\tf \omm 157}&{\tf \omm 208}&{\tf \omm 174}&{\tf  23}&{\tf 0.9}&{\tf 0.8}&{\tf      80}&{\tf 0.27}&{\tf 0.12}&{\tf  102~~}&{\ttf BB58,\,HP735} & {\ttf (2); (5)} \\
{\tf 110}&{\tf ?HVC\,236.7\omp 49.8\omp 078}&{\tf 10 25.4}&{\tf \omp 06 42}&{\tf       78}&{\tf \omm  41}&{\tf \omm  67}&{\tf  37}&{\tf 1.3}&{\tf 1.3}&{\tf       0}&{\tf 0.19}&{\tf 0.14}&{\tf  293~~}&{\ttf WvW47,\,BB59} & {\ttf (1); (5)} \\
{\tf 111}&{\tf ?HVC\,241.0\omp 53.4\omp 089}&{\tf 10 43.5}&{\tf \omp 06 41}&{\tf       89}&{\tf \omm  26}&{\tf \omm  57}&{\tf  60}&{\tf 1.3}&{\tf 1.0}&{\tf      50}&{\tf 0.15}&{\tf 0.18}&{\tf  281~~}&{\ttf WvW34,\,BB60} & {\ttf (1); (5)} \\
{\tf 112}&{\tf :HVC\,260.1\omp 47.8\omp 217}&{\tf 11 02.3}&{\tf \omm 05 46}&{\tf      217}&{\tf       72}&{\tf       26}&{\tf  49}&{\tf 0.8}&{\tf 0.8}&{\tf       0}&{\tf 0.18}&{\tf 0.10}&{\tf   33~~}&{\ttf HP974} & \\
{\tf 113}&{\tf ?HVC\,284.0\omm 84.0\omm 174}&{\tf 01 00.7}&{\tf \omm 32 47}&{\tf \omm 174}&{\tf \omm 196}&{\tf \omm 167}&{\tf  29}&{\tf 1.3}&{\tf 1.2}&{\tf       0}&{\tf 0.35}&{\tf 0.19}&{\tf  351~~}&{\ttf WvW561,\,BB64,\,HP1417} & {\ttf (2); (5)} \\
{\tf 114}&{\tf ?HVC\,340.1\omp 22.5\omm 108}&{\tf 15 29.6}&{\tf \omm 28 43}&{\tf \omm 108}&{\tf \omm 177}&{\tf \omm 257}&{\tf  32}&{\tf 1.3}&{\tf 0.8}&{\tf      60}&{\tf 0.36}&{\tf 0.23}&{\tf  280~~}&{\ttf BB65} & {\ttf (2); (5)} \\
{\tf 115}&{\tf :HVC\,357.5\omp 05.6\omp 268}&{\tf 17 17.9}&{\tf \omm 27 59}&{\tf      268}&{\tf      259}&{\tf      192}&{\tf  45}&{\tf 0.4}&{\tf 0.4}&{\tf       0}&{\tf 0.24}&{\tf 0.13}&{\tf   81~~}&{\ttf HP1966} &   \\
{\tf 116}&{\tf CHVC\,357.5\omp 12.4\omm 181}&{\tf 16 53.4}&{\tf \omm 23 58}&{\tf \omm 181}&{\tf \omm 190}&{\tf \omm 260}&{\tf  51}&{\tf 0.8}&{\tf 0.8}&{\tf       0}&{\tf 0.27}&{\tf 0.14}&{\tf   97~~}&{\ttf BB66,\,HP1974} &   \\
\hline
\end{tabular}
\end{table*}

Each of the individual isolated objects retrieved from the LDS by the
search algorithm and listed in Table~\ref{table:CHVC} is illustrated in
Figs. \ref{fig:stamps1} -- \ref{fig:stamps5} by an image of integrated
\hi emission, paired with a spectrum.  The images show \hi integrated
over the velocity range of emission which is considered part of the
cloud; the gray--scale intensities are given by the color bar in units
of K\,\kms.  The associated spectrum refers to the direction in the
$0\fdg5 \times 0\fdg5$ grid nearest to the peak of the integrated
emission.  Inspection of the Digital Sky Survey in the region of each
of these images did not show a clear optical counterpart for any of the
listed objects.

\begin{figure*}
\centering
\caption{Images of integrated \hi emission, paired with a
representative spectrum for all of the fully and partially isolated
objects (CHVCs, :HVCs and ?HVCs) retrieved from the LDS by the search
algorithm, confirmed in independent data, and cataloged in
Table~\ref{table:CHVC}.  The images show \hi integrated over the entire
velocity range of emission which is considered part of the cloud; the
gray--scale intensities are given by the color bar in units of K\,\kms.
Contours are drawn for \NH\ = 1.5, 3, 4.5 and 6$\times$10$^{18}$
cm$^{-2}$. The associated spectrum refers to the direction in the
$0.5^\circ \times 0.5^\circ$ grid nearest to the peak of the integrated
emission.  The data for the images and for the spectra were extracted
from the LDS, not from the independent confirming material.
}\label{fig:stamps1}
\end{figure*}

\begin{figure*}
\centering
\caption{Images of integrated \hi emission, paired with a
representative spectrum for all of the fully and partially
isolated objects (CHVCs, :HVCs and ?HVCs)
retrieved from the LDS by the search algorithm as in Fig.\ref{fig:stamps1}
}\label{fig:stamps2}
\end{figure*}

\begin{figure*}
\centering
\caption{Images of integrated \hi emission, paired with a
representative spectrum for all of the fully and partially
isolated objects (CHVCs, :HVCs and ?HVCs)
retrieved from the LDS by the search algorithm as in Fig.\ref{fig:stamps1}
}\label{fig:stamps3}
\end{figure*}

\begin{figure*}
\centering
\caption{Images of integrated \hi emission, paired with a
representative spectrum for all of the fully and partially
isolated objects (CHVCs, :HVCs and ?HVCs)
retrieved from the LDS by the search algorithm as in Fig.\ref{fig:stamps1}
}\label{fig:stamps4}
\end{figure*}

\begin{figure*}
\centering
\caption{Images of integrated \hi emission, paired with a
representative spectrum for all of the fully and partially
isolated objects (CHVCs, :HVCs and ?HVCs)
retrieved from the LDS by the search algorithm as in Fig.\ref{fig:stamps1}
}\label{fig:stamps5}
\end{figure*}

The arrangement on the sky of the isolated objects listed in
Table~\ref{table:CHVC} is shown superposed on velocity--integrated sky
images in Fig.~\ref{fig:skynegneg}, for the velocity range $-450 <
V_{\rm LSR} < -150$ \kms; in Fig.  \ref{fig:skyneg}, for the range
$-150 < V_{\rm LSR}< -90$ \kms; and in Fig. \ref{fig:skypos}, for the
range $+90 < V_{\rm LSR} < +150$ \kms.  We briefly comment on the sky
deployment below; De Heij et al.  (\cite{deheij02}) discuss it more
fully in conjunction with the southern--hemisphere catalog.

\begin{figure*}
\caption{Distribution of CHVCs found in the velocity range $-450 <
V_{\rm LSR} < -150$ \kms~across the northern hemisphere (left) and the
portion of the southern hemisphere at $\delta > -30\deg$ accessed by
the LDS (right).  Open diamonds indicate the locations of individual
CHVCs.  The gray shadings indicate the total emission in this velocity
range, with the color bar giving the scale in units of K\,\kms.
Circles of constant declination are draw for $\delta =
0^\circ$,~$\pm30^\circ$, and~$\pm60^\circ$.  The scale--bar shows units
of K\,\kms.  Bright, isolated clouds in the images which are not
centered within a CHVC symbol are nearby galaxies.  Due to the large
range of integration, some individual compact clouds -- of narrow width
or otherwise of low total flux, do not clearly show up.  Confirmation
has proven, however, that they are real.}\label{fig:skynegneg}
\end{figure*}

\begin{figure*}
\caption{Like Fig.~\ref{fig:skynegneg}, but with the velocity
integration ranging from $V_{\rm LSR} = -150\rm\;km\;s^{-1}$ to 
$V_{\rm LSR} = -90\rm\;km\;s^{-1}$.}\label{fig:skyneg}
\end{figure*}

\begin{figure*}
\caption{Like Fig.~\ref{fig:skynegneg}, but with the velocity
integration ranging from $V_{\rm LSR} = +90\rm\;km\;s^{-1}$ to 
$V_{\rm LSR} = +350\rm\;km\;s^{-1}$.}\label{fig:skypos}
\end{figure*}

The distribution of isolated object sizes and linewidths are shown in
Fig.~\ref{fig:sizefwhm}. Although the only limit
on angular size we have imposed is the $10\deg \times 10\deg$ dimension
of our initial column density image of each candidate, the distribution
is strongly peaked with a median at $1\deg$ FWHM and does not extend
beyond $2\fdg2$.  Sharply bounded anomalous--velocity objects
apparently can be well described with the term: Compact High Velocity
Clouds. The distribution of velocity FWHM is also strongly peaked at
about 25 \kms, although the distribution does extend out to 100~\kms.

\begin{figure*}
\caption{Histograms of the angular size and velocity width of all of
the fully and partially isolated objects (CHVCs, :HVCs and ?HVCs)
retrieved from the LDS by the search algorithm, confirmed in
independent data, and cataloged in Table~\ref{table:CHVC}. Although all
HVCs less than 10$\deg$ in diameter where considered, the median CHVC
is only 1$\deg$ and the maximum $2\fdg2$.}\label{fig:sizefwhm}
\end{figure*}

\subsection{Differences between this catalog and that of BB99}\label{subsec:BB}

The BB99 compilation was based on the same data as were used in the
preparation of Table~\ref{table:HVC} and \ref{table:CHVC}, but on
selection criteria which were somewhat different than those of the
algorithm described here, and so there are several differences between
the Table~\ref{table:CHVC} catalog and that of BB99.  The
Table~\ref{table:CHVC} catalog includes objects not found in the BB99
list, because it was extended to lower flux levels since new confirming
data were planned; conversely, a number of objects listed by BB99 were
not included in the current work.  The differences are explained as
follows:

\begin{description}
\item[] {\em BB19, BB45, BB47, BB49, BB50, BB53, BB59 and BB60:} These
objects all have a $V_{\rm DEV}$ less than the~$70\rm\,km\,s^{-1}$
limit conservatively imposed for our search.  The first of these
objects has been identified as a nearby LSB galaxy, Cep I, by Burton et
al. (\cite{burton99}) and is not included in Table~\ref{table:CHVC}.
But the other seven objects are included in the table to allow direct
comparison with the rest of the CHVC sample. The high background \NH\
levels that are inevitable with such a low deviation velocity, result
in none of these objects passing the more stringent isolation criterion
employed here, even though this background appears smooth on angular
scales of 5--10$^\circ$ and is conceivable unrelated to the object in
question. They have been given the ?HVC designation to indicate the
uncertainty in classification.
\item[] {\em BB12, BB14, BB18, BB35, BB37, BB42, BB61, BB62 and BB63:}
Although these clouds were classified as isolated in the BB99 study,
they do not satisfy the more stringent critera for isolation applied
here and are therefore not included in the listing.  The BB99 crition
of isolation was based on the 50\% contour of peak \NH; here we used
the 1.5$\times$10$^{18}$ cm$^{-2}$ contour. 
\item[] {\em BB17, BB20, BB24, BB28, BB33, and BB34:} These 6~objects
are included in Table~\ref{table:CHVC}, but have been classified here
as :HVCs, based on the more restrictive criteria of the current
work. There is some indication for extended emission in the environment
of these sources which might be associated.
\item[] {\em BB2, BB4, BB26, BB46, BB48, BB55, BB56, BB58, BB64 and
BB65:} These 10~objects are also included in Table~\ref{table:CHVC},
but have been classified here as ?HVCs. In all these cases the
background \NH\ level exceeds our limit of 1.5$\times$10$^{18}$
cm$^{-2}$ but appears smooth on angular scales of 5--10$^\circ$ and is
conceivable unrelated to the object in question.
\end{description}

Thus of the 65 compact anomalous--velocity clouds listed by BB99
(excluding the nearby galaxy Cep 1), the criteria applied here have
resulted in the identification of 54; of these, 31 have retained the
designation CHVC, whereas 6 have been assigned the designation :HVC and
17 have been labeled ?HVC. Some examples of reclassified objects are
shown in Fig.~\ref{fig:reclass}.  Of the total of 116 compact objects
tabulated, 32 do not appear in the Wakker \& van Woerden
(\cite{wakker91}) catalog; of these 32 objects, 17 are classified in
Table~\ref{table:CHVC} as CHVCs.

\begin{figure}
\caption{ Examples of objects which were identified as CHVCs by BB99,
but which have been re--classified following the more stringent
criteria described in this paper, which measures the isolation of a
feature using the \NH\ contour at a fixed level of 1.5$\times$10$^{18}$
cm$^{-2}$, rather than at the level of 50\% of the peak in each object
as used by BB99. The features shown in the upper two panels were
reclassified as :HVCs, thus as possible members of the class of compact
objects; but the features shown in the two lower panels are not
isolated according to the new criterion, and were reclassified as
HVCs. The cross in the lower left of each panel indicates the angular
extent of a true degree on the sky. The contours are drawn at 1.5, 3,
4.5 and 6$\times$10$^{18}$ cm$^{-2}$; the intensity scale is indicated
by the color bar, in units of K\,\kms. }
\label{fig:reclass}
\end{figure}

\subsection{Differences in the zone of overlapping declinations 
between this catalog and the HIPASS catalog}\label{subsec:HIPASS}

Putman et al. (\cite{putman02}) have applied the search algorithm
described here to the HIPASS data, resulting in a catalog of southern
compact, isolated objects.  Because both the LDS catalog of CHVCs given
here in Table~\ref{table:CHVC} and the HIPASS catalog will be used
together in an all--sky study of the kinematic and spatial properties
of CHVCs, a comparison between them in the zone of overlap is
interesting.  The two surveys overlap in the declination range $-28\deg
\leq \delta \leq +2\deg$, but were carried out with different
observational parameters. The RMS noise figure is 10~mK in the HIPASS
material, for a channel 26 \kms~wide and a FWHP beam of $15^\prime$;
the corresponding RMS value in the LDS is 70~mK, for a channel width of
1.03 \kms~and a FWHP beam of $36^\prime$.  After smoothing both surveys
to the same spectral resolution of 26 \kms, the 3$\sigma$ limiting
column density (for emission filling each beam) is 0.47 and
0.64$\times$10$^{18}$ cm$^{-2}$ in the HIPASS and LDS respectively.
Thus, while the sensitivity to well-resolved sources is comparable, the
point source sensitivity of HIPASS is greater by about a factor of 3.
On the other hand, the band-pass calibration of the HIPASS data relies
on reference spectra which are deemed empty of \hi emission that are
off-set by only a few degrees on the sky. Even with the {\sc MINMED5}
method of baseline determination employed by Putman et al., there is
significant filtering of extended emission, which complicates the
assessment of object isolation down to a low column density limit.

The differences in point-source sensitivity on the one hand and
sensitivity to very extended structures on the other, results in a
larger number of faint source detections in HIPASS, but also in a
different designation for some of the brighter clouds which the surveys
have in common. Comparison of the results derived by applying the
search algorithm to both surveys allows assessment of the robustness of
the selection criteria and of the completeness of the LDS catalog.

Table~\ref{table:comp} lists the number of anomalous--velocity features
classified as either CHVCs or :HVCs and found in the indicated
peak--temperature bins (referred to the HIPASS temperature calibration
scale) in both the HIPASS and LDS material in the zone of overlapping
declinations. The data indicate the influence of survey sensitivity on
the number of objects found.  Above a peak temperature of~0.45~K, the
LDS results are as complete as those based on the HIPASS; between
$T_{\rm peak}=0.20$ K and 0.45 K, the LDS results recovered 83\% of the
clouds found in the HIPASS data.  The completeness of the LDS catalog
drops rapidly at lower values of the peak temperature: HIPASS clouds
less bright than 0.20 K are almost completely absent from the LDS
catalog.  The incompleteness at low peak temperatures will be more
important for the smaller CHVCs than for the somewhat less compact
:HVCs. Some of the smaller objects have a total column density which is
difficult to distinquish from the LDS spectral noise (see
Subsec.~\ref{subsec:select}).  Their FWHM angular sizes are less
than~$25^\prime$ and they may be centered as much as $10^\prime$ from
the nearest LDS telescope pointing.  The small sizes of the CHVCs and
the less--than--Nyquist sampling interval of the LDS can conspire to
result in an observation with a lower signal--to--noise ratio in the
LDS than in the HIPASS.
 
The differences in sensitivity not only influence the number of sources
that are detected, but also the way in which they are classified.
Within the zone of overlapping declinations, 33 CHVCs were found in the
HIPASS material with a peak temperature above~0.20~K.  Of these
33~sources, 19 were also found in the LDS.  The appearance of these 19
sources in the LDS led 7 of them to be classified as CHVCs, 7 as :HVCs,
and the remaining 5 as HVCs. The difference in assignment is primarily
due to the differences in the properties of the two sets of data. The
possibility of a different designation is greater for the fainter
sources.  For the eight sources with a HIPASS peak temperature
above~0.35~K, four have a different LDS designation, whereas there is
only agreement for three of the eleven sources which are fainter
than~$0.35$~K.  Evidently the differing sensitivities to compact and
extended structures of the LDS and HIPASS do not allow for consistent
classification of the weakest objects.  Figure~\ref{fig:LDSHIPASS}
shows, as an example, the HIPASS data for a cloud which is seen in
projection against an extended filament: the LDS data were unable to
detect the weak emission of the filament, and as a consequence the
search algorithm applied to the LDS returned a CHVC designation.  This
example demonstrates that the classification of clouds depends on the
sensitivity of the survey.

\begin{figure}
\caption{Velocity--integrated images of two objects which were
retrieved by the search algorithm both from the HIPASS material as
reported by Putman et al. (\cite{putman02}) and from the LDS material
discussed here, but which were classified differently by the search
algorithm because of the differing observational parameters of the two
surveys. The images show the HIPASS data on the left after spatial
smoothing to 36 arcmin FWHM; the LDS, on the
right.  The angular orientation and the velocity range of integration
are the same for both of the paired images.  For the object represented
in the upper pair, a classification of CHVC followed from the HIPASS
material, but a classification of :HVC from the LDS data.  The object
represented in the lower pair of images was classified as a HVC from
the HIPASS material but as CHVC from the LDS.  The cross in the lower
left of each panel indicates the angular extent of a true degree on the
sky.  The contours are drawn at \NH\ = 1.5, 3, 4.5 and
6$\times$10$^{18}$ cm$^{-2}$; the intensity scale is indicated by the
color bar, in units of K\,\kms. }
\label{fig:LDSHIPASS}
\end{figure}

If the HIPASS and LDS catalogs are compared or if they are used
together, for example to investigate the all--sky properties of the
compact--object ensemble, then due attention should be given to the
higher expected detection rate in the southern material, and to its
greater sensitivity to the most compact features.  A straightforward
merger of the two catalogs would neglect the higher rate of detections
in the southern hemisphere and the possibly different designation of a
limited number of the clouds.  To correct for the higher detection rate
in the south, Table~\ref{table:comp} can be used to roughly estimate
the likelihood that a given cloud which is observed in the HIPASS data
also will be observed in the LDS.

\subsection{Completeness and homogeneity of the LDS sample of CHVCs}

Before discussing some aspects of the LDS sample of CHVCs, a few 
additional remarks concerning completeness and homogeneity of the sample
are in order. These parameters are influenced by the non--Nyquist sampling
of the LDS and its finite sensitivity, 

Due to the finite sensitivity of the LDS, we will have missed clouds
with peak temperatures or column densities below threshold values.
After converting the HIPASS peak temperatures in Table
~\ref{table:comp} to the LDS temperature scale, that table can be used
to estimate the number of undetected features.  A comparison of the
objects with detections both in the HIPASS and in the LDS listings
shows that the average of the ratio between the HIPASS and LDS peak
temperatures equals~1.5. This difference can be understood in terms of
the differing angular sampling intervals and resolutions of the two
surveys.

\begin{table}
\caption{Number of clouds classified as CHVCs and :HVCs in the overlap
zone of the HIPASS and LDS surveys, listed according to the peak
brightness temperatures of the clouds.   The data from the two surveys differ
sufficiently that the search algorithm described here may have returned a
different classification from the HIPASS material than from the LDS.  The
peak temperatures refer to the HIPASS temperature calibration. The
statistics show that the LDS catalog of compact anomalous--velocity objects 
is likely to be complete for clouds with $T_{\rm peak} > 0.3$\,K, within
the velocity range constrained by the bandwidth of the survey and by the
adapted deviation velocity. }
\label{table:comp}
\begin{center}
\begin{tabular}{|c|r|c|}
\hline
$T_{ \rm peak}$ & 
    \multicolumn{2}{c|}{\# CHVCs $+$ :HVCs} \\
\cline{2-3} 
({\sc k}) & 
    \multicolumn{1}{|c|}{ ~~\sc hipass~~}  & \multicolumn{1}{c|}{\sc lds} 
\\
\hline
$0.10 \ldots 0.15$  &  28~~~~   &   0     \\
$0.15 \ldots 0.20$  &  26~~~~   &   4     \\
$0.20 \ldots 0.25$  &  17~~~~   &   7     \\
$0.25 \ldots 0.30$  &   8~~~~   &   5     \\
$0.30 \ldots 0.35$  &  10~~~~   &   9     \\
$0.35 \ldots 0.40$  &   8~~~~   &   7     \\
$0.40 \ldots 0.45$  &   2~~~~   &   1     \\
$0.45 \ldots 0.50$  &   3~~~~   &   3     \\
$0.50 \ldots 0.55$  &   1~~~~   &   1     \\
$0.55 \ldots 0.60$  &   3~~~~   &   3     \\
\hline
\end{tabular}
\end{center}
\end{table}

The validity of Fig.~\ref{fig:completeness} as an indication of
completeness is suggested by the detection rate for known external
galaxies which appear in the LDS. Hartmann \& Burton
(\cite{hartmann97}) list all of the galaxies that are cataloged in LEDA
and detected in the LDS.  Of the known external galaxies with $|V_{\rm
DEV}| > 70$ \kms~and $V_{\rm LSR} < 350$ \kms, all those with a peak
\hi brightness temperature greater than~0.13~K were found by the search
algorithm.  Of the~12 galaxies with $0.09\,{\rm K} \le T_{\rm peak} <
0.12\,{\rm K}$, 42\% were found. The temperatures and therefore
sensitivities were measured at the grid points of the survey; point
sources that are not located at a grid point will have been observed
with a reduced sensitivity, which depends on the telescope beam and the
distance to the nearest grid point.  The LDS sampled the sky on a
$0\fdg5$ by $0\fdg5$ lattice; the sensitivity away from the grid points
falls off as a Gaussian with a FWHP of~36$^\prime$.  The results, shown
in Fig.~\ref{fig:completeness}, lead us to consider the LDS catalog to
be 100\% complete for point sources with $T_{\rm peak} \ge 0.3$\,K.
The solid curve follows the trend as indicated by the HIPASS catalog.

\begin{figure}
\caption{Degree of completeness expected from the application of the
search algorithm for point sources with given temperature. The
full--drawn line shows the expected completeness based on comparison
between the results of our search and a catalog of nearby galaxies
extracted from LEDA and the LDS.  The histogram shows, for objects
lying at declinations in the region of overlap of the HIPASS and LDS
survey, the fraction of CHVCs which are listed both in the HIPASS
catalog of Putman et al. (\cite{putman02}) and in the LDS catalog of
Table~\ref{table:CHVC}.  The histogram follows Table~\ref{table:comp}, after
transforming the HIPASS temperatures to the LDS scale.  The average
ratio of the HIPASS and LDS peak temperature for a given cloud is not
unity (but equals 1.5), due to the differences in the spectral and
spatial resolutions of the two surveys.  }\label{fig:completeness}
\end{figure}

Objects with \hi properties like the CHVCs will be missed if their
emission is superposed on emission from intermediate-- or
high--velocity complexes, or from our Galaxy.  Independent of their
exact origin, it is likely that some of these clouds will coincide with
extended emission of a different type.  Related to this problem is the
selection based on the deviation velocity of the clouds.  We mentioned
above that 12\% of the BB99 sample were excluded from this study by
this selection criterion.  Before we can properly investigate the exact
influence of the obscurations or exclusion of some parts of the LDS, we
have to model the intrinsic distribution of the CHVCs and then assess
how they would appear in the data set.  We  investigate this issue
in our analysis of the all-sky distribution of CHVCs (De Heij et
al. \cite{deheij02}). 

Although the part of the LDS that was searched only covered the sky
within the velocity interval $V_{\rm LSR} = -450$ \kms~to $+350$ \kms,
there are indications that we do not miss many (if any) clouds because
of the velocity interval.  The high--velocity feature with the most
extreme negative velocity yet found is HVC\,$110.6-07.0-466$, discoverd by
Hulsbosch (\cite{hulsbosch78}) and subject to substantial subsequent
observation as referenced in Table~\ref{table:CHVC}.  The Wakker \& van Woerden
tabulation, which relied on survey data covering the range $-900$ \kms~to
$+750$ \kms, found no high--velocity cloud at a more negative velocity. 
The HIPASS search reported by Putman et al. (\cite{putman02}) sought
anomalous--velocity emission over the range $-700 < V{\rm LSR} < +1000$
\kms.  Of the 194 HIPASS CHVCs cataloged by Putman et al., ten have
$V_{\rm LSR}<-300$, but the most extreme negative velocity is $-353$
\kms.  This CHVC, namely CHVC\,$125.1-66.4-353$ occurs, not
surprisingly, in the quadrant where the northern data shows a
preference for extreme negative velocities.  

The north/south kinematic asymmetries are a well--known property of the
anomalous--velocity \hi. In terms of the Local Group deployment
discussed by Blitz et al. (\cite{blitz99}) and by BB99, the most
extreme negative velocities would be found in the general region of the
barycenter of the Local Group, whereas the most extreme positive LSR
velocities, which would be more modest in amplitude than the extreme
negative velocities, would be found in the general region of the
anti--barycenter of the Local Group. The LDS does not reach low enough
declinations to embrace the anti--barycenter region, although the
feature in Table~\ref{table:CHVC} with the highest positive velocity,
:HVC\,$357.5+05.6+268$ is well removed from the direction of the
barycenter\footnote{Although none of the compact \hi clouds listed in
Table~\ref{table:CHVC} showed an obvious optical counterpart in our
perusal of the DSS, it is not yet firmly ruled out that the listing
might be contaminated by an obscured nearby galaxy, as yet undetected
at optical or infrared wavelengths.  Two of the features listed in our
catalog appear particularly suspicious in this regard, namely
:HVC\,$357.5+05.6+268$ and CHVC\,$004.4+05.7+202$. Both are located
along sight lines traversing the Galactic Bulge, and thus lie in
directions of exceptionally high optical obscuration, and both deviate
strongly from the kinematic patterns defined by the other members of
the ensemble.  This deviation is particularly evident in
Fig.~\ref{fig:helvel}, a figure of the sort frequently used to
establish membership in the Local Group.}.  Of the 194 CHVCs in the
HIPASS catalog, only 7 have $V_{\rm LSR}$ greater than $+300$ \kms, and
only one has a velocity greater than $350$ \kms, namely
CHVC\,$258.2-23.9+359$.  All of the seven CHVCs with substantial
positive velocities lie deep in the third longitude quadrant, or in the
fourth; the mean longitude of these seven CHVCs is $278\deg$, $185\deg$
removed from the longitude appropriate for the solar apex motion in the
Local Group Standard of Rest reference frame as determined by
Karachentsev \& Makorov (\cite{karachentsev96}), and whichthus roughly
corresponds with the direction of the Local Group anti--barycenter. The
Wakker \& van Woerden compilation lists no high--velocity feature with
a more positive velocity than that of their HVC\,$305.0-10.0+312$, also
in the general direction preferentially represented by positive
velocites.

\begin{figure}
\caption{Heliocentric velocities of nearby galaxies and of the ensemble
of compact high--velocity clouds, plotted as a function of
$\cos(\theta$), where $\theta$ is the angle between the object and the
location of the apex of the solar motion relative to the center of mass
of the Local Group. CHVCs are shown as triangles, and the
less--constrained :HVC and ?HVC objects as diamonds and squares,
respectively; the Local Group galaxies, from the tabulation of Mateo
 (\cite{mateo98}), are plotted as stars. The solid line represents the
solar motion of $V_\odot = 316$ \kms~toward $l=93\deg$, $b=-4\deg$ as
determined by Karachentsev \& Markarov (\cite{karachentsev96}).  The
dashed lines show the envelope one standard deviation ($\pm 60$ \kms)
about the $V_{\rm HEL}, \cos(\theta)$ relation, pertaining for galaxies
considered firmly established as members of the Local Group. }
\label{fig:helvel}
\end{figure}

In view of these detection statistics, we consider it unlikely that the
velocity range of the LDS has caused a significant number of features
to be missed at the declinations observed.  In other words, the true
velocity extent, as well as the non--zero mean in the LSR frame, of the
anomalous--velocity ensemble appear well represented by the extrema of
$-466$ \kms~and $+359$ \kms.

\section{Discussion}
\label{sect:discussion}

\subsection{What defines a CHVC?}

The concept of a distinct class of compact, isolated high--velocity
clouds has emerged from the visual inspection of large area images of
good sensitivity and spatial sampling like the LDS in the north and the
HIPASS in the south. There appears to be a class of high-contrast
features which are at best only marginally resolved with half degree
angular resolution and that can not easily be distinguished from the
\hi signature of an external galaxy. The \hi signature of an external
galaxy in the LDS (see Figure 15 of Hartmann \& Burton \cite{hartmann97}) is
a moderately high peak column density of a few times 10$^{19}$ cm$^{-2}$
or more (averaged over the 36 arcmin beam), an \hi FWHM linewidth
varying between 20 and 375 \kms commensurate with the galaxy mass and
inclination, and a {\it sharply bounded angular extent}, such that emission
at column densities above a few times 10$^{18}$ cm$^{-2}$ is confined to
less than about 2$^\circ$ diameter.

This last fact is of particular physical relevance, since the
precipitous decline in \hi column density seen at the edges of nearby
galaxies can be understood as arising from photo--ionization due to the
intergalactic radiation field. Detailed studies of individual systems
(NGC3198 by Maloney \cite{maloney93}, M33 by Corbelli \& Salpeter
\cite{corbelli93}) show that an exponential decline in neutral column
density, with a scale--length of about 1~kpc sets in below a critical
column density of about 2$\times$10$^{19}$ cm$^{-2}$.  This neutral
column density is comparable to the Warm Neutral Medium layer (with a
temperature of about 8000~K) required to provide sufficient shielding
from UV and soft X-ray radiation such that condensation of Cool Neutral
Medium clumps (with temperature of about 100~K) can take place
(e.g. Wolfire et al. \cite{wolfire95a}) for ambient thermal pressures
comparable to, or less than, those found in the solar
neighbourhood. The picture that emerges is a nested structure of CNM
cores, shielded by WNM cocoons and surrounded by a Warm Ionized Medium
halo. For nearby galaxies the WNM cocoon at the edge of the gaseous
disk has an observed exponential scale-length of about 1~kpc (Maloney
\cite{maloney93}; Corbelli \& Salpeter \cite{corbelli93}). The
exponential scale-length of the partially ionized WNM cocoon is
determined by both the total column density distribution and the
incident ionizing spectrum. For smooth distributions of total column
density and a wide range of power--law spectral indices, Corbelli \&
Salpeter (\cite{corbelli93}) find neutral scale--lengths of about
1~kpc. The reason that this transition zone is so much more extended
than in a classical \hii region is found in the wide range of photon
energies, and hence penetrating depths, of the ionizing spectrum.

It is not   obvious that all of these considerations
need apply to the population of high--velocity clouds. If the HVCs are 
near the Galactic disk, then they will be subject to severe tidal
distortions of their intrinsic gas distributions and be exposed to
variable ionizing radiation levels depending on local
circumstances. Due to changing physical conditions it is conceivable
that thermal and pressure equilibrium may not be achieved. The
requirements for achieving such equilibrium have been considered by
Wolfire et al. (\cite{wolfire95b}), who conclude that even for heights
above the Galactic plane of only 3~kpc, thermal equilibrium should be
marginally achieved within the WNM and easily within the CNM at an
infall velocity as high as 150~\kms. At larger distances the
requirements that the thermal timescale be shorter than the time to
experience significant pressure variations are even more easily
satisfied. It thus seems reasonable to expect that high--velocity
clouds might also be described by CNM clumps within a WNM shielding
cocoon surrounded by WIM halos. If such objects occur at distances
greater than about 60~kpc, tidal effects would be less disruptive, the
UV radiation field would be more nearly isotropic, and an \hi
concentration of about 2~kpc diameter would have an angular size of
less than 2$^\circ$.

\subsection{Application of a digital filter to the LDS}

Motivated by the distinctive visual appearance of sharply bounded \hi
peaks in the LDS data, we have considered whether a simple digital
filtering might not serve to isolate a subset of these features from
the general \hi emission of the Galaxy and the HVC complexes, and in so
doing offer independent confirmation of the results achieved with the 
procedure outlined in \S\ref{sect:algorithm}.   We considered a
two--dimensional discrete derivative convolving kernel, consisting of a
delta function at the origin together with a negative ring of unit integral
at a radius of 1$^\circ$. After evaluating the convolution of each channel
map with this kernel, we normalize the result by the original intensity,
wherever this exceeds some minimum significance:
\begin{equation}
F(x,y) =  \frac{1}{I} \frac{dI}{dr}
\end{equation}
for $I > n\sigma$. As an illustration, this filter was applied to an
Aitoff-projected cube of the LDS data centered at
$(l,b)=(123\deg,0\deg)$ after a velocity smoothing to 25 \kms~FWHM, and
restricted to intensities exceeding 10$\sigma$, corresponding to 0.14~K
brightness. The peak filtered response was then determined along each
spectrum; this is illustrated as the greyscale in Figure
\ref{fig:skydis} for values exceeding 0.85 in units of the normalized
derivative. Overlaid on the peak filter values are square symbols at
the positions of 59 cataloged external galaxies.  Of these 59 nearby
\hi--emitting galaxies, 57 are listed in Table 4 of Hartmann \& Burton
(\cite{hartmann97}); the remaining two are Dwingeloo 1, discovered by
Kraan--Korteweg et al. (\cite{kraan94}), and Cepheus 1, discovered by
Burton et al. (\cite{burton99}).  Open circles in this figure show the
positions of the 116 CHVCs which we catalog here; crosses show the
positions of 72 CHVC candidates which were not confirmed in subsequent
observations, and triangles show the positions of 84 CHVC candidates
for which no confirming observations have yet been obtained.

\begin{figure*}
\centering
\caption{Distribution on the sky of the compact anomalous--velocity
objects (CHVCs) found by application of the search algorithm and
selection criteria to the Leiden/Dwingeloo Survey. The background
greyscale is the output of a spatial derivative filter to the LDS data
(matched to the median cloud linewidth and angular size) where this
exceeds 0.85 in units of the normalized derivative. Overlaid are square
symbols at the positions of 59 cataloged external galaxies, open
circles for the 116 CHVCs which we catalog here; crosses at the
positions of 72 CHVC candidates which were not confirmed in subsequent
observations, and triangles at the positions of 84 CHVC candidates for
which no confirming observations have yet been obtained.  The CHVCs are
not strongly clumped in the northern sky; and in particular show no
concentration toward the extended HVC complexes or Magellanic Stream
but do show some concentration toward the region near
(l,b)=(25$\deg$,$-$30$\deg$) which has previously been called the
Galactic Center Negative velocity population. The solid curve indicates
the nominal $\delta = -30^\circ$ cut-off of the LDS coverage.  }
\label{fig:skydis}
\end{figure*}

It is striking how the vast majority of both normal Galactic and HVC
complex emission has been eliminated by this simple filtering. About
half of the peaks returned by the normalized derivative filter have an
over--plotted symbol corresponding either to a cataloged external galaxy or
to a CHVC. The 18 cataloged galaxies and 39 CHVCs which do not have a strong
filter response can all be understood in terms of a poorly matched
velocity width (since objects well--matched with a 25 \kms~FWHM were
selected in this case) or a very low peak brightness (that is, below the
cut--off of 0.14~K). In addition to the filter peaks with over--plotted
symbols there are a similar number, about 140 peaks, exceeding 0.85 in the
normalized spatial derivative without an overplotted identification. Closer
inspection of these additional filter peaks reveals that they all
correspond to sub--structure within more extended HVC complexes. These
are quite distinct from the CHVCs in that they are not are not sharply
bounded at a column density of 1.5$\times$10$^{18}$ cm$^{-2}$, but
instead are connected (at least in projected position--velocity space) to
more extended HVC complexes at a higher column density.

\subsection{Brief remarks on the spatial and kinematic deployment of the
northern CHVCs}

In a separate paper (De Heij et al. \cite{deheij02}), we  analyse
the LDS CHVC sample merged with the sample identified by Putman et
al. (\cite{putman02}) in the southern hemisphere HIPASS material.  Here
we briefly note the global properties of the compact objects
found in the LDS.

Figure \ref{fig:skydis} shows the distribution on the sky of the
compact, isolated objects cataloged in Table~\ref{table:CHVC}. The
objects are distributed rather uniformly across the northern sky. In
particular, there is no clear sign of a preference in the sky
distribution either for the Galactic disk, or for most of the known
high--velocity cloud complexes, or the Magellanic Stream. One exception
to this general conclusion is an apparent concentration centered near
(l,b)=(25,$-$30), a region which Wakker \& Van Woerden
(\cite{wakker91}) identify with their Galactic Center Negative Velocity
Population of HVCs. This region appears to be particuarly rich in
compact objects statisfying very stringent requirements for isolation
in column density. For comparison, the distribution of all nearby
galaxies detected in the LDS data is also plotted in the figure.

Figure~\ref{fig:veldis} shows the velocities of all of the compact
clouds cataloged in Table~\ref{table:CHVC}, calculated for three
different reference frames and plotted as function of the Galactic
longitude.  The velocity dispersion of the sample decreases from 174
\kms, to 101 \kms, and then to 97 \kms, in going from the reference
frame of the Local Standard of Rest, to that of the Galactic Standard
of Rest, and then to that of the Local Group Standard of Rest.  The
average velocity changes from $-191$ \kms, to $-127$ \kms, and then to
$-117$ \kms, for the three different reference frames,
respectively. Although the differences between the Galactic and Local
Group system are not very large, the values are lower in the Local
Group coordinate system. Inclusion of the data of the southern
hemisphere in the analysis has to indicate if the small difference is
significant or not.  The difference of 10 \kms~between the velocity
dispersion in the LGSR reference frame found by BB99 and that found
here is probably due to the differences in the samples used, with the
BB99 one being smaller.

\begin{figure}
\caption{Kinematic deployment of the compact--object sample plotted as
a function of the galactic longitude for three different kinematic
reference frames, compared with the kinematic distribution of Local
Group galaxies.  The CHVCs are shown as triangles; the
less--constrained :HVC and ?HVC objects as diamonds and squares,
respectively.  The kinematic distribution of the Local Group galaxies
tabulated by Mateo (\cite{mateo98}) is traced by the filled circles.
There is a decrease in the average velocity of the compact--object
ensemble, as well as in the velocity dispersion, when progressing from
the Local Standard of Rest system (upper panel), to the Galactic
Standard of Rest system (middle panel), to the Local Group Standard of
Rest system (lower panel). }
\label{fig:veldis}
\end{figure}

\subsection{Summary and Conclusions}

An automated procedure has been developed to extract
anomalous--velocity clouds from \hi survey material.  We have applied
the algorithm to the Leiden/Dwingeloo survey and catalog the properties
of a total of 917 HVC features with a deviation velocity in excess of
70~\kms. Since the algorithm requires the existence of a local maximum
in position--velocity which is distinct from the bulk of Galactic
emission, the catalog can not be usefully extended to include the
intermediate velocity clouds which are strongly blended with the
Galaxy.  We have searched our HVC catalog for all isolated clouds,
defined by having a lowest significant column density contour
(3$\sigma\sim1.5\times$10$^{18}$ cm$^{-2}$) which is (1) closed, with
its greatest radial extent less than $10\deg$ by $10\deg$; and (2) not
elongated in the direction of any nearby extended emission. A total of
116 objects have been tabulated which at least partially satisfy these
criteria. Independent confirmation is available for all of these
clouds, some of which appear in only single spectra in the LDS.  Of the
116 clouds, 54 had been identified as CHVCs by Braun \& Burton
(\cite{braun99}), some others had been detected in earlier work
referenced in the catalog but are confirmed as isolated by the data and
analysis presented here, and others are reported here for the first
time.

Although objects as large as $10\deg$ were permitted by our
selection criteria, the resulting distribution is strongly peaked at a
median value of $1\deg$ FWHM and has a maximum observed diameter of
only $2.\fdg2$. Isolated HVCs are observed to be relatively
compact. Conversely, although the well known HVC complexes exhibit
a wealth of small-scale structure that is comparable in angular scale
($\sim1\deg$) these structures are {\it not isolated}. The local maxima
within the HVC complexes are surrounded by extended emission with
column densities in the range $5-20\times$10$^{18}$ cm$^{-2}$. 

The significance of high galactic latitude \hi features which are
isolated in column density down to a level as low as
$1.5\times$10$^{18}$ cm$^{-2}$, is that this is about an order of
magnitude lower than the critical column density identified at the
edges of nearby galaxies (Maloney \cite{maloney93}, Corbelli \&
Salpeter \cite{corbelli93}), $\sim$2$\times$10$^{19}$ cm$^{-2}$, where
the ionized fraction is thought to increase dramatically due to the
extragalactic radiation field. Unless very contrived geometries are
invoked of some unseen population of high column density absorbers,
these objects will need to provide their own shielding to ionizing
radiation. (This point has also been made by Hoffman et
al. \cite{hoffman01}.) Self-consistent calculations of the ionization
balance in the shielding layer exposed to a power--law extragalactic
ionizing photon field yield a typical \hi exponential scale--length of
1~kpc (Corbelli \& Salpeter \cite{corbelli93}).  The small median
angular size of the CHVCs, of about $1^\circ$ FWHM, might then imply
substantial distances, greater than about 120~kpc.

The kinematic and spatial deployment of the enlarged sample shows that
both the Galactic Standard of Rest and the Local Group Standard of Rest
frames substantially lower the velocity dispersion of the
population. This also lends support to the hypothesis that the CHVCs
are a distant component of the high--velocity cloud phenomenon, at
substantial distances and with a net in--fall towards the Galaxy or
the Local Group barycenter.

A more complete analysis of the kinematic and spatial deployment of the
all--sky CHVC ensemble incoporates both the LDS sample and the results
of a comparable search by Putman et al. (\cite{putman02}) in the HIPASS
data, and is reported separately by De Heij et al. (\cite{deheij02}).

\begin{acknowledgements} 
The Westerbork Synthesis Radio Telescope is operated by the Netherlands
Foundation for Research in Astronomy, under contract with the Netherlands
Organization for Scientific Research. 

\end{acknowledgements}


\begin{thebibliography}{} 

\bibitem[1985]{arp85}
	Arp H., 1985, AJ, 90, 1012
\bibitem[1985]{bajaja85}
    Bajaja E., Cappa de Nicolau C.\,E., Cersosimo J.\,C., Loiseau N.,
    Martin M.\,C., Morras R., Olano C.\,A., P\"oppel W.\,G.\,L., 1985,
    ApJS, 58, 143
\bibitem[1987]{bajaja87}
Bajaja E., Morras R., P\"oppel W.\,G.\,L., 1987,
Pub. Astr. Inst. Czech. Ac. Sci., 69, 237
\bibitem[1975]{baker75}
	Baker P.\,L., Burton W.\,B., 1975, ApJ, 198, 281
\bibitem[2001]{barnes01}
    Barnes D.\,G., Staveley--Smith L., de Blok W.\,J.\,G., Oosterloo
T., Stewart I.\,M., Wright A.\,E., Banks G.\,D., Bjathal R., et al.,
2001, MNRAS, 322, 486
\bibitem[1998]{binney98}
	Binney J., Merrifield M., 1998, Galactic Astronomy,
	Princeton University Press
\bibitem[1999]{blitz99}
	Blitz L., Spergel D.\,N., Teuben P.\,J., Hartmann D., Burton W.\,B.,
1999,
	ApJ, 514, 818
\bibitem[1999]{braun99}	
Braun R., Burton W.\,B., 1999, A\&A, 341, 437 (BB99)
\bibitem[2000]{braun00}
	Braun R., Burton W.\,B., 2000, A\&A, 354, 853 
\bibitem[2001]{bruns01}
 Br\"uns C., Kerp J., Pagels A., 2001, A\&A, 371, 816
\bibitem[2001]{burton01a}
Burton W.\,B., Braun R., Chengalur J.\,N., 2001, A\&A, 369, 616
\bibitem[2001]{burton01b}
Burton W.\,B., Braun R., Chengalur J.\,N., 2001, A\&A, 375, 227
\bibitem[1999]{burton99}
Burton W.\,B., Braun R., Walterbos R.\,A.\,M., Hoopes C.\,G., 1999, AJ,
117, 194
\bibitem[1979]{cohen79}
Cohen R.\,J., Mirabel I.\,F., 1979, MNRAS, 186, 217
\bibitem[1993]{corbelli93}
Corbelli E., Salpeter E.\,E., 1993, ApJ, 419, 104
\bibitem[1975]{davies75}
Davies R.\,D., 1975, MNRAS, 170, 45P
\bibitem[2002]{deheij02}
De Heij V., Braun R., Burton W.\,B., 2002, A\&A, submitted
\bibitem[1976]{eichler76}
	Eichler D., 1976, ApJ, 208, 694
\bibitem[1976]{einasto76}
Einasto J., Haud U., J\^oeveer M., Kaasik A., 1976, MNRAS, 177, 357
\bibitem[1981]{giovanelli81}
Giovanelli R., 1981, AJ, 86, 1468
\bibitem[1994]{hartmann94}
	Hartmann D., 1994, Ph. D. Thesis, University of Leiden
\bibitem[1996]{hartmann96}
	Hartmann D., Kalberla P.\,M.\,W., Burton W.\,B., Mebold U., 1996,
	A\&AS, 119, 115
\bibitem[1997]{hartmann97}
	Hartmann D., Burton W.\,B., 1997,
	Atlas of Galactic Neutral Hydrogen, Cambridge University Press
\bibitem[1992]{henning92}
Henning P.\,A., 1992, ApJS, 78, 365
\bibitem[2001]{hoffman01}
Hoffman G.\,L., Salpeter E.\,E., Pocceschi M.\,G., 2001, ApJ, submitted
\bibitem[1978]{hulsbosch78}
Hulsbosch A.\,N.\,M., 1978, A\&A, 66, L5
\bibitem[1988]{hulsbosch88}
	Hulsbosch A.\,N.\,M., Wakker B.\,P., 1988, A\&AS, 75, 191
\bibitem[1996]{karachentsev96}
Karachentsev I.\,D., Makarov D.\,A., 1996, AJ, 111, 794
\bibitem[1994]{kraan94}
Kraan--Korteweg R.\,C., Loan A.\,J., Burton W.\,B., Lahav O., Ferguson
H.\,C., Henning P.\,A., Lynden--Bell D., 1994, Nature, 372, 77
\bibitem[1993]{maloney93}
Maloney P., 1993, ApJ, 414, 41
\bibitem[1998]{mateo98}
 Mateo M., 1998, ARA\&A, 36, 435
\bibitem[1981]{mirabel81}
Mirabel I.\,F., 1981, ApJ, 247, 97
\bibitem[1979]{mirabel79}
Mirabel I.\,F., Cohen R.\,J., 1981, MNRAS, 188, 219
\bibitem[1963]{muller63}
	Muller C.\,A., Oort J.\,H., Raimond E., 1963,
	C. R. Acad. Sci. Paris, 257, 1661
\bibitem[1966]{oort66}
	Oort J.\,H., 1966, Bull. Astr. Inst. Netherlands, 18, 421
\bibitem[1970]{oort70}
	Oort J.\,H., 1970, A\&A, 7, 381
\bibitem[1981]{oort81}
	Oort J.\,H., 1981, A\&A, 94, 359
\bibitem[1999]{putman99}
Putman M., Gibson B.K., 1999, PASA, 16, 70
\bibitem[2002]{putman02}
    Putman M., de Heij V., Staveley--Smith L., Braun, R., Freeman K.\,C.,
Gibson B.\,K., Burton W.\,B., Barnes D.\,G., et al., 2002, AJ,
in press, January issue
\bibitem[1974]{saraber74}
Saraber M.\,J.\,M., Shane W.\,W., 1974, A\&A, 30, 365
\bibitem[1990]{stutzki90}
    Stutzki J., G\"usten R., 1990, ApJ, 356, 513
\bibitem[1998]{thilker98}
    Thilker D.\,A., Braun R., Walterbos R.\,A.\,M., 1998, A\&A, 332, 429
\bibitem[1999]{vanwoerden99}
	van Woerden H., Schwarz U.\,J., Peletier R.\,F., Wakker B.\,P.,
	Kalberla P.\,M.\,W., 1999, Nature, 400, 138
\bibitem[1975]{verschuur75}
	Verschuur G.\,L., 1975, ARA\&A, 13, 257
\bibitem[1999]{voskes99}
 Voskes T., Burton W.\,B., 1999, in: ASP Conf. Ser. 168, New Perspectives on
the Interstellar Medium, A.\,R. Taylor, T.\,L. Landecker, \& G. Joncas
(eds.), p. 375
\bibitem[1990]{wakker90}
	Wakker B.\,P., 1990, Ph.\,D. Thesis, University of Groningen
\bibitem[2001]{wakker01}
Wakker B.\,P., 2001, ApJS, 136, 463
\bibitem[1991]{wakker91}
	Wakker B.\,P., van Woerden H., 1991, A\&A, 250, 509
\bibitem[1997]{wakker97}
	Wakker B.\,P., van Woerden H., 1997, ARA\&A, 35, 217
\bibitem[1999]{wakker99}
	Wakker B.\,P., van Woerden H., Gibson B.\,K., 1999,
	in: ASP Conf. Ser. 166, Stromlo Workshop on High--Velocity Clouds,
	B.\,K. Gibson \& M.\,E. Putman (eds.), p.\ 311
\bibitem[1994]{williams94}
    Williams J.\,P., de Geus E.\,J., Blitz L., 1994,
    ApJ, 428, 693
\bibitem[1995a]{wolfire95a}
Wolfire M.\,G., Hollenbach D., McKee C.\,F., Bakes E.\,L.\,O.,
1995, ApJ, 443, 152
\bibitem[1995b]{wolfire95b}
Wolfire M.\,G., McKee C.\,F., Hollenbach D., Tielens A.\,G.\,G.\,M.,
1995, ApJ, 453, 673
\bibitem[1979]{wright79}
Wright M.\,C.\,H., 1979, ApJ, 233, 35
\end{thebibliography}
\end{document}